\documentclass[10pt,twocolumn,letterpaper]{article}

\usepackage{cvpr}              
\newcommand{\red}[1]{{\color{red}#1}}
\newcommand{\blue}[1]{{\color{blue}#1}}

\newcommand{\Rmnum}[1]{\MakeUppercase{\romannumeral #1}}

\usepackage[table]{xcolor}
\usepackage{colortbl}
\definecolor{tableHeadGray}{gray}{.9}
\usepackage{multirow}
\usepackage{float}
\usepackage{adjustbox}
\usepackage{etoolbox}
\usepackage{dsfont}
\newcommand{\pub}[1]{{\color{gray}{\tiny{[{#1}]\!}}}}

\usepackage{tabularray}
\usepackage{arydshln}
\usepackage{bm}
\usepackage{amsmath}

\usepackage{xcolor}

\def\eg{\textit{e.g.}}

\definecolor{cvprblue}{rgb}{0.21,0.49,0.74}
\usepackage[pagebackref,breaklinks,colorlinks,allcolors=cvprblue]{hyperref}
\usepackage{xcolor}
\usepackage{multirow}
\usepackage{booktabs}
\usepackage{graphicx} 
\usepackage[table]{xcolor} 
\usepackage{makecell}
\usepackage{pifont}
\usepackage{amssymb}
\usepackage{caption}
\usepackage{algorithm}    
\usepackage{algpseudocode}

\usepackage{pgfplots}
\pgfplotsset{compat=1.17}
\definecolor{Gray}{gray}{0.95} 
\newcommand{\cmark}{\textcolor{green!50!black}{\checkmark}} 
\newcommand{\xmark}{\textcolor{red!50!black}{$\times$}}     

\def\paperID{} 
\def\confName{CVPR}
\def\confYear{2026}

\makeatletter
\renewcommand*{\@fnsymbol}[1]{\ensuremath{\ifcase#1\or *\or \dagger\or \ddagger\or
    \mathsection\or \mathparagraph\or \|\or **\or \dagger\dagger
    \or \ddagger\ddagger \else\@ctrerr\fi}}
\makeatother

\title{Learning Latent Transmission and Glare Maps for Lens Veiling Glare Removal}
\author{
Xiaolong Qian$^{1,*}$ \quad Qi Jiang$^{1,*}$ \quad Lei Sun$^{2,\dagger}$ \quad Zongxi Yu$^{1}$ \quad Kailun Yang$^{3}$ \quad Peixuan Wu$^{1}$\\Jiacheng Zhou$^{1}$ \quad Yao Gao$^{1}$ \quad Yaoguang Ma$^{1}$ \quad Ming-Hsuan Yang$^{4,5}$ \quad Kaiwei Wang$^{1,\dagger}$\\
\normalsize
$^{1}$Zhejiang University \quad $^{2}$INSAIT, Sofia University ``St. Kliment Ohridski''\\
\normalsize 
$^{3}$Hunan University \quad $^{4}$University of California, Merced \quad $^{5}$Google DeepMind
}

\begin{document}

\twocolumn[{%
\renewcommand\twocolumn[1][]{#1}%
\maketitle
\begin{center}
    \centering
    \captionsetup{type=figure}
    \includegraphics[width=0.99\linewidth]{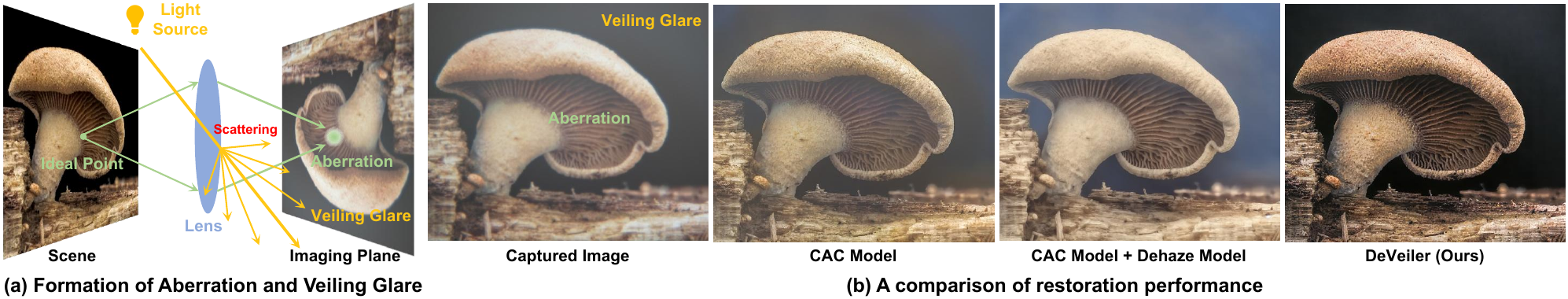}
    \vskip-1.5ex
    \captionof{figure}{Simplified optical systems suffer from residual aberrations and veiling glare (a), caused by design-induced blur and stray-light scattering from non-ideal surfaces and coatings. (b) Existing methods fail under this compound degradation: a Computational Aberration Correction (CAC) model~\cite{liang2021swinir} retrained on aberration-only data fails on unseen veiling glare, while a cascaded state-of-the-art dehazing model~\cite{wang2025learning} introduces inconsistent artifacts. Our DeVeiler restores a clean image by jointly correcting the compound degradations.}
    \label{fig:teaser}
\end{center}
}]

{  \renewcommand{\thefootnote}{$^\ast$}
  \footnotetext{Equal contribution. $^\dagger$Corresponding authors.}
}

\begin{abstract}
Beyond the commonly recognized optical aberrations, the imaging performance of simplified optical systems—including single-lens and metalens designs—is often further degraded by veiling glare caused by stray-light scattering from non-ideal optical surfaces and coatings, particularly in complex real-world environments. This compound degradation undermines traditional lens aberration correction yet remains underexplored. A major challenge is that conventional scattering models (\eg, for dehazing) fail to fit veiling glare due to its spatial-varying and depth-independent nature. Consequently, paired high-quality data are difficult to prepare via simulation, hindering application of data-driven veiling glare removal models. To this end, we propose VeilGen, a generative model that learns to simulate veiling glare by estimating its underlying optical transmission and glare maps in an unsupervised manner from target images, regularized by Stable Diffusion (SD)-based priors. VeilGen enables paired dataset generation with realistic compound degradation of optical aberrations and veiling glare, while also providing the estimated latent optical transmission and glare maps to guide the veiling glare removal process. We further introduce DeVeiler, a restoration network trained with a reversibility constraint, which utilizes the predicted latent maps to guide an inverse process of the learned scattering model. Extensive experiments on challenging simplified optical systems demonstrate that our approach delivers superior restoration quality and physical fidelity compared with existing methods. These suggest that VeilGen reliably synthesizes realistic veiling glare, and its learned latent maps effectively guide the restoration process in DeVeiler. All code and datasets will be publicly released at \url{https://github.com/XiaolongQian/DeVeiler}.
\end{abstract}    
\section{Introduction}
\label{sec:intro}

Driven by the demand for compact and high-performance imaging systems in applications like Augmented Reality (AR/VR)~\cite{li2021meta,chen2024hybrid} and mobile photography~\cite{chen2021extreme_quality,chen_mobile_2023}, simplified optical systems such as single-lens~\cite{qian2025towards} and metalens~\cite{tseng2021neural,zhang2025degradation} are becoming increasingly important. 
However, simplified optics inevitably introduce aberrations due to design trade-offs, while low-cost and structural constraints cause surface imperfections and non-ideal coatings, resulting in diffuse stray-light scattering manifested as \textit{veiling glare} (see Fig.~\ref{fig:teaser}(a)).
The resulting veiling glare manifests as a \textit{widespread and diffuse veil} that reduces image contrast even under normal lighting conditions, in contrast to structured artifacts such as lens flare or ghosting, which typically occur when bright sources are in or near the field of view~\cite{kotp2023toward_flare_free}.
Consequently, residual aberration and veiling glare frequently coexist, forming a \textit{compound degradation} that severely limits image quality in simplified systems.
While Computational Aberration Correction (CAC)~\cite{heide2013high,chen2021extreme_quality,chen2021optical,li2021universal,jiang2024minimalist,jiang2024flexible,gong2024physics,chen2025physics,li2025optical_attention} can mitigate spatial-varying aberrations, it cannot recover the contrast loss induced by diffuse scattering.
Moreover, hardware suppression is highly constrained: applying advanced coatings to such lenses is manufacturing-prohibitive, and baffles fail against \textit{in-field} stray light.
These factors highlight the urgent need for a computational framework capable of modeling and restoring this coupled optical degradation.
Addressing this compound degradation is non-trivial, as the absence of realistic training data remains a key bottleneck. 
Physically-accurate simulation of veiling glare demands both complete optomechanical models and computationally prohibitive non-sequential ray-tracing, making large-scale data generation impractical~\cite{raskar2008glare,hullin2011physically}.
This data scarcity constrains existing paradigms. 
Classical approaches to veiling glare removal, such as deconvolution-based methods~\cite{faulkner1989veiling,talvala2007veiling} or image decomposition~\cite{zhang2018single}, often rely on restrictive assumptions like spatial invariance and tend to fail in low-texture scenes.
Modern learning-based methods that rely on simplified 2D synthesis models~\cite{shoshin2021veiling} suffer from a significant domain gap.
Moreover, methods designed for visually similar degradations are fundamentally incompatible. 
Flare removal algorithms~\cite{dai2022flare7k,dai2024flare7k++,tsai2025lightsout,FlareX_lishenqu} are ineffective as they target structured artifacts, not the diffuse nature of veiling glare. 
Dehazing methods~\cite{he2010single,wu2023ridcp,wang2024ucl,wang2025learning} are similarly unsuitable, as their atmospheric scattering models are incompatible with the \textit{depth-independent} physics of glare originating within the lens assembly.

\begin{figure*}[!t]
  \centering
  \includegraphics[width=0.90\linewidth]{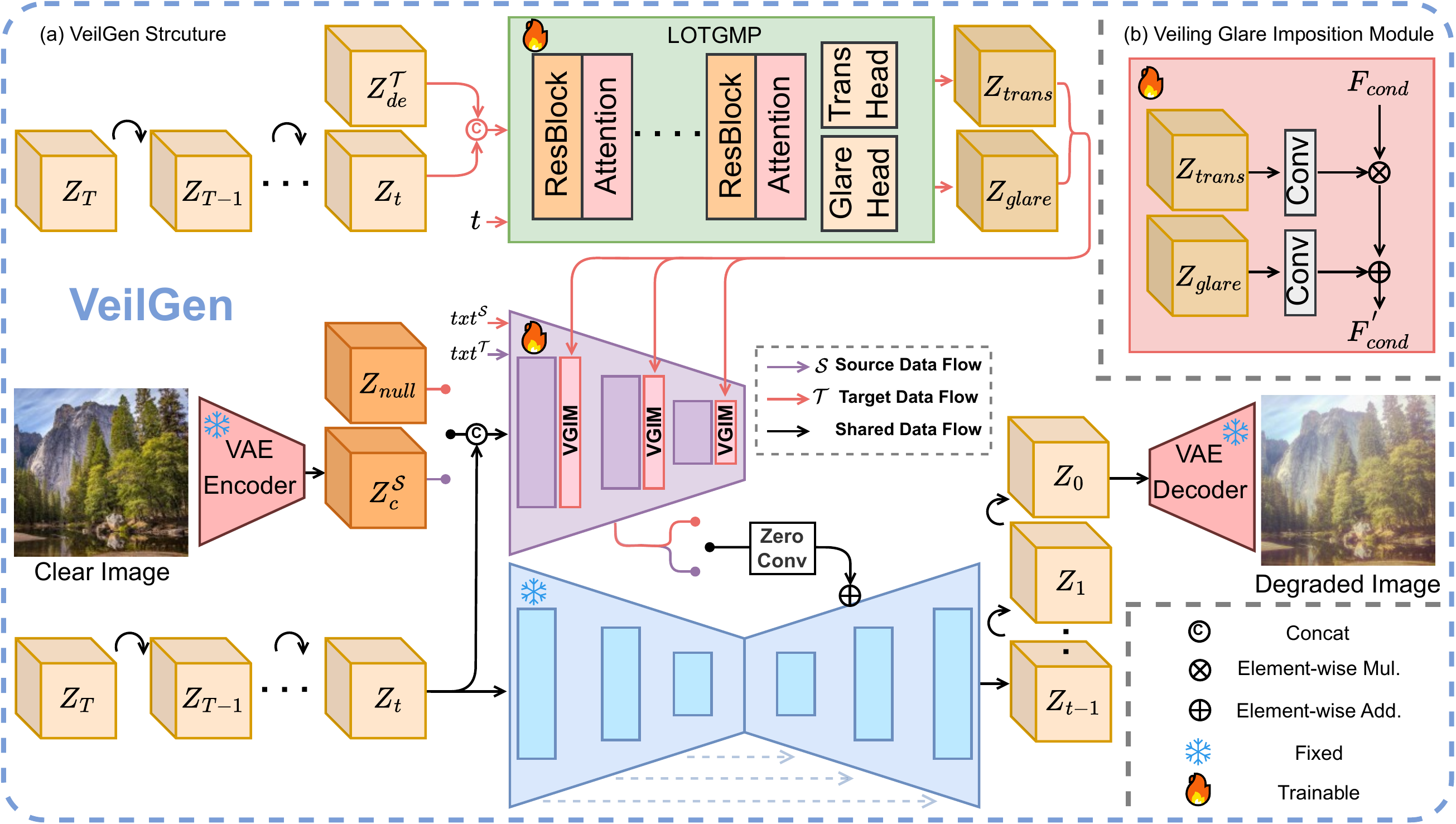}
  \vskip-1.5ex
  \caption{\textbf{Overall architecture of the proposed VeilGen.} In Stage~\Rmnum{1}, VeilGen is trained to synthesize compound degradations by using a Latent Optical Transmission and Glare Map Predictor (LOTGMP) to estimate latent maps. These maps then guide the diffusion process via the Veiling Glare Imposition Module (VGIM). 
  $Z_{de}^{\mathcal{T}}$ is obtained by encoding target degraded images via the frozen VAE.
  $Z_{t}$ denotes the noisy latent representation at timestep $t$ of the forward diffusion process. $Z_{null}$ represents an all-zero latent representation. $txt^{\mathcal{S}/\mathcal{T}}$ denotes the text prompts for the source and target domains, respectively.}
  \label{fig:veilgen}
  \vspace{-1.5em}
\end{figure*}

To address this data challenge, recent Stable Diffusion (SD)-based generative models~\cite{yang2023synthesizing,peng2024towards,wang2025learning} have been explored, yet they often operate as ``black-boxes'', lacking physical grounding. 
In contrast, we introduce VeilGen, a physics-informed SD-based model that embeds glare formation principles directly into the generative process (Fig.~\ref{fig:veilgen}). 
In our setting, paired data are only available for aberration correction, while unpaired degraded images from the simplified systems exhibit veiling glare on top of residual aberrations, with physical components that are difficult to measure precisely.
To bridge this gap, VeilGen incorporates a Latent Optical Transmission and Glare Map Predictor (LOTGMP) that estimates the two key latent components, the transmission and glare maps, within the diffusion process.
The latent maps are injected via the Veiling Glare Imposition Module (VGIM) to modulate features, enabling VeilGen to synthesize realistic paired data (see \S\ref{sec:stage_1}).
Building upon our generative foundation, we propose DeVeiler, a restoration network trained with a reversibility constraint (Fig.~\ref{fig:deveiler}(b)). 
Unlike pure blind-learning approaches, which produce an averaged, suboptimal removal for spatial-varying veiling glare, DeVeiler is structurally guided to learn an inverse mapping of the degradation.
However, directly using VeilGen during restoration training is impractical because its multi-step diffusion sampling is computationally expensive.
We therefore distill the generative process into a lightweight forward model that preserves its behavior and latent-map conditioning, and use it to supervise the training process of DeVeiler (see \S\ref{sec:stage_2}).
Its core component, the Veiling Glare Compensation Module (VGCM), uses internally estimated latent maps to modulate image features, thereby precisely reversing the forward glare imposition process (see \S\ref{sec:stage_3}). 
This allows DeVeiler to remove glare based on its underlying physical causes instead of spurious statistical correlations, achieving superior restoration results, as shown in Fig.~\ref{fig:teaser}(b).
While our ultimate goal is addressing the \textit{compound degradation} of simplified optics, our core technical novelty focuses specifically on \textit{veiling glare removal}. 
Since aberrations are sufficiently mitigated via foundational source-domain supervision, the under-constrained veiling glare remains the critical bottleneck (detailed in the Supplement). 
Validated on two challenging lens prototypes, our framework achieves state-of-the-art restoration by effectively utilizing shared latent maps to explicitly model and reverse this physical glare.
In a nutshell, our contributions are summarized as follows:
\begin{itemize}
\item We propose VeilGen, a novel physics-informed generative model. It features a Latent Optical Transmission and Glare Map Predictor (LOTGMP) to estimate physical maps and a Veiling Glare Imposition Module (VGIM) that uses these maps to guide a diffusion process for synthesizing realistic compound optical degradations.
\item We introduce DeVeiler, a restoration network trained with a reversibility constraint to approximate the inverse of the learned degradation. Supervision from a lightweight distilled forward model enforces consistency between the forward VGIM and the inverse feature modulation in the Veiling Glare Compensation Module (VGCM).
\item Extensive experiments demonstrate that our method achieves state-of-the-art performance in the challenging task of joint aberration and veiling glare removal.
\end{itemize}

\section{Related Work}
\label{sec:Related_Work}

\noindent\textbf{Computational Aberration Correction.}
Computational Aberration Correction (CAC) employs algorithms to compensate residual optical aberrations. 
Recent learning-based approaches~\cite{peng2019learned,chen2021extreme_quality,chen2021optical,jiang2024minimalist,luo2024correcting,zhang2025degradation,mandal2025enabling} now outperform traditional optimization-based methods~\cite{wiener1949extrapolation,fish1995blind,schuler2012blind,2015blind}.
However, their reliance on simulated data causes a persistent sim-to-real gap and limits generalization across lenses. 
Efforts toward universal CAC frameworks~\cite{li2021universal,gong2024physics,jiang2024flexible} improve cross-lens adaptability using diverse lens libraries, yet they primarily target aberration-induced blur. 
The concurrent veiling glare remains underexplored, constraining performance under compound degradations.

\noindent\textbf{Veiling Glare Removal.}
Research on glare-related artifacts generally falls into two categories.
Lens flare removal~\cite{wu2021how_to_train,dai2022flare7k,dai2023nighttime,dai2024flare7k++,vasluianu2024sfnet,tsai2025lightsout,zhu2025pbfg,FlareX_lishenqu}, targets structured, localized artifacts such as bright spots and streaks.
In contrast, veiling glare manifests as a diffuse degradation that reduces image contrast, making flare-oriented approaches ineffective.
Prior work relied on deconvolution using the Glare Spread Function (GSF)~\cite{talvala2007veiling} or on image decomposition~\cite{zhang2018single}, yet these methods impose restrictive assumptions, often require multiple captures, or fail in low texture scenes.
More recent learning-based techniques are limited by the lack of realistic paired data, as existing synthesis strategies~\cite{shoshin2021veiling} remain overly simplified.
Moreover, previous studies typically handle veiling glare and aberration as independent degradations, whereas our framework jointly addresses both within a unified computational model.

\noindent\textbf{Domain Adaptation in Low-level Vision.}
While Domain Adaptation (DA) mitigates domain shifts in low-level vision~\cite{wei2021unsupervised,wang2021unsupervised,li2019semi,chen2021psd}, generic approaches struggle with compound degradations coupling optical aberrations and veiling glare.
This motivates specialized DA methods adapting pre-trained models to novel lenses~\cite{jiang2025qdmr,jiang2024flexible}.
Although target-domain synthesis via SD-based generative models~\cite{yang2023synthesizing,peng2024towards,wang2025learning} shows promise, black-box approaches remain suboptimal for well-defined physical optical degradations.
To this end, our work introduces a physics-informed generative model that explicitly estimates latent maps for veiling glare.
These maps serve dual roles: conditioning an SD-based generator to produce realistic degradations and guiding the restoration network to better remove veiling glare.

\vspace{-0.5em}
\section{Methodology}
\label{sec:method}
\subsection{Problem Formulation}
This work aims to restore a latent clean image $I_{c}$ from a degraded observation $I_{de}$ captured by a simplified optical system.
The degradations central to this work are twofold: (i) spatial-varying aberration blur and (ii) veiling glare.
We model the forward degradation process in two sequential steps.
Following the widely adopted patch-wise convolution model~\cite{li2021universal,yang2023aberration,luo2024correcting,yang2025efficient}, the aberration image patch $I_{ab}^p$ is obtained by convolving each clean patch $I_{c}^p$ with its corresponding Point Spread Function (PSF) $K^p$, representing local aberration characteristics.
This aberration patch $I_{ab}^p$ is then sequentially degraded by veiling glare according to scattering model, formulated as a combination of attenuation and additive light~\cite{zhang2018single,shoshin2021veiling}: each patch $I_{ab}^p$ is modulated by a transmission map $T^p$ and combined with a glare map $I_g^p$, where $T^p$ denotes local contrast attenuation and $I_g^p$ represents the veiling glare.
For each color channel (omitted for brevity), the overall degradation model is expressed as:
\vspace{-1.0mm}
\begin{equation}
I_{de}^p = \underbrace{(I_{c}^p \otimes K^p)}_{I_{ab}^p} \cdot T^p + I_{g}^p.
\label{eq:Img Forward}
\end{equation}

The restoration of $I_{c}^p$ from $I_{de}^p$ thus constitutes a highly ill-posed blind inverse problem.
The numerous unknowns in the degradation model make optimization-based methods with generic priors ineffective for this problem.
This motivates a data-driven approach capable of learning the inverse mapping directly from data.
However, its success fundamentally depends on the availability of high-quality paired training data $(I_{de}, I_{c})$, which is rarely accessible in practice.
Accordingly, our work focuses on two key objectives: (1) generating paired data that accurately captures the target degradation characteristics and (2) designing a restoration network that effectively leverages it.
Formally, the overall objective is to learn optimal network parameters $\theta$ via
\begin{equation}
\theta^* = \arg\min_{\theta} \mathcal{L}(f_{\theta}(I_{de}), I_{c}),
\label{eq:objective}
\end{equation}
where $\mathcal{L}$ denotes an appropriate loss function.

\subsection{Overall Framework}
\begin{figure}[!t]
  \centering
  \includegraphics[width=0.99\linewidth]{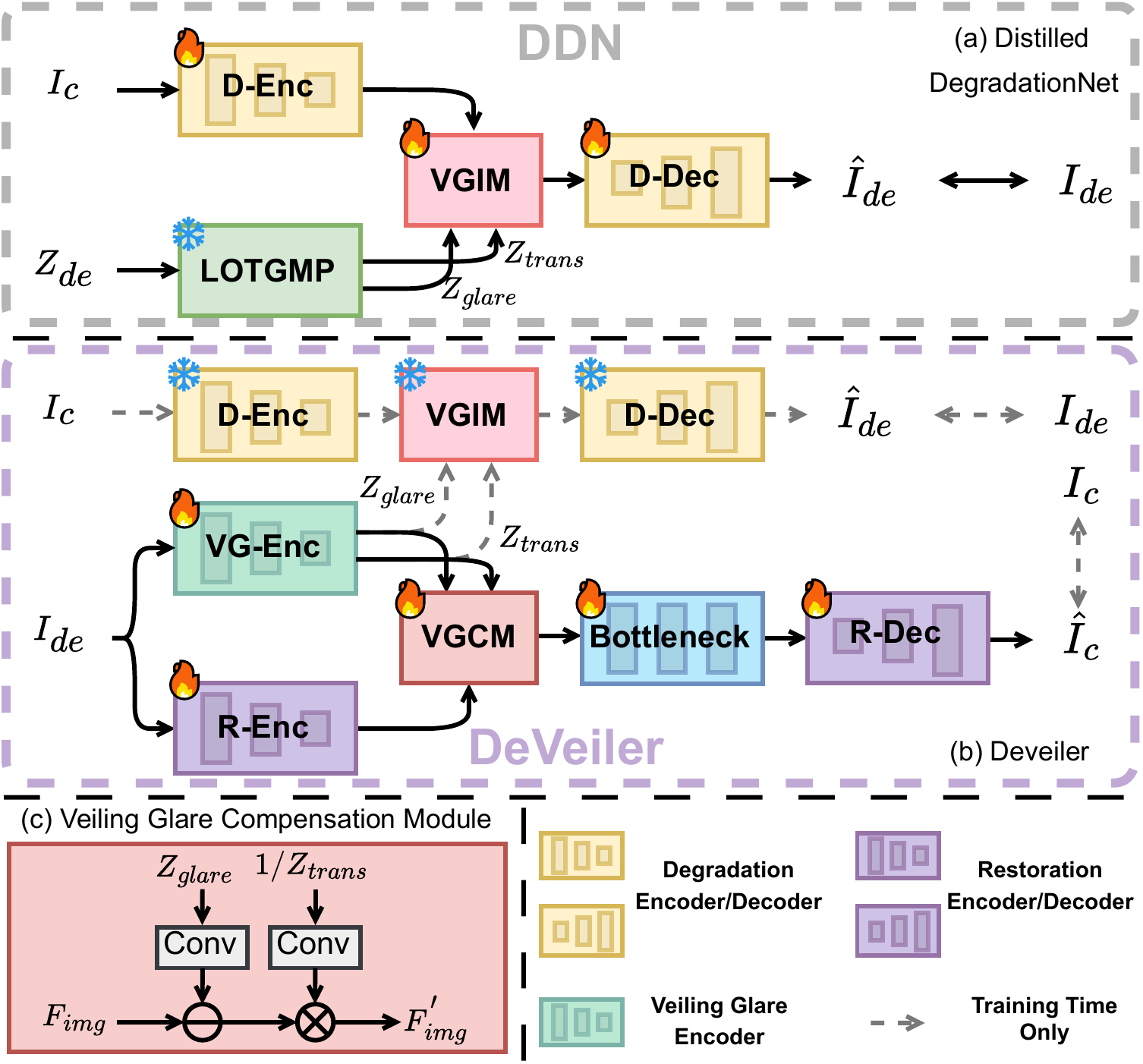}
  \vskip-1.5ex
    \caption{\textbf{The distillation and restoration networks.} (a) The Distilled Degradation Net (DDN), trained in Stage~\Rmnum{2}, models the forward degradation, using VGIM to apply the veiling glare prior. (b) The DeVeiler, trained in Stage~\Rmnum{3}, is designed to reverse this process: it first removes the veiling glare using the VGCM, and then feeds the intermediate result into its main bottleneck to correct the aberrations.}
  \label{fig:deveiler}
  \vspace{-1em}
\end{figure}

Our framework addresses compound degradations through a three-stage, physics-informed pipeline that unifies data synthesis and restoration via shared latent priors.
In Stage~\Rmnum{1}, VeilGen, an SD-based generator, synthesizes realistic compound degradations. Its LOTGMP predicts latent transmission and glare maps, which guide the Veiling Glare Imposition Module (VGIM) to modulate image features and produce realistic degradations (Fig.~\ref{fig:veilgen}).
In Stage~\Rmnum{2}, VeilGen is distilled into a lightweight Distilled Degradation Net (DDN) that preserves its degradation behavior while remaining efficient enough to supervise restoration training (Fig.~\ref{fig:deveiler}(a)). 
In Stage~\Rmnum{3}, the restoration network, DeVeiler, is trained to invert the forward degradation. It achieves this by leveraging the predicted latent maps via its Veiling Glare Compensation Module (VGCM) to invert the scattering process of the VGIM (Fig.~\ref{fig:deveiler}(b)).

\subsection{Stage \Rmnum{1}: Realistic Degradation Generation}
\label{sec:stage_1}
To overcome data scarcity, we introduce VeilGen, an SD-based generative model that synthesizes realistic paired data (Fig.~\ref{fig:veilgen}(a)).
Built upon DiffBIR~\cite{lin2024diffbir} with a fixed Stable Diffusion backbone and IRControlNet, VeilGen adapts the standard denoising paradigm~\cite{ho2020denoising} and incorporates a novel mechanism for modeling veiling glare.

\noindent\textbf{Physics-informed Hybrid Training.} 
To bridge a source domain $\mathcal{S}$ with paired, glare-free data and a target domain $\mathcal{T}$ with unpaired compound degradations, we propose a physics-informed hybrid training paradigm inspired by recent work in domain adaptation~\cite{wang2025learning}. 
Since the spatial-varying transmission and glare maps governing veiling glare are difficult to obtain directly, the diffusion model alone cannot capture the underlying physics.
To address this, we introduce the Latent Optical Transmission and Glare Map Predictor (LOTGMP), which infers these latent maps during denoising, providing physically meaningful guidance to the generative process.
Unlike conventional end-to-end degradation learning, VeilGen steers generation using unified latent maps $c_{vg} = (z_{trans}, z_{glare})$:
\begin{equation}
\label{eq:cvg_definition}
c_{vg} =
\begin{cases}
    (\mathbf{1}, \mathbf{0}) & \text{if in source domain,} \\
    \text{LOTGMP}(z_t, z_{de}^\mathcal{T}, t) & \text{otherwise.}
\end{cases}
\end{equation}

For the source domain, we use fixed maps ($\mathbf{1}, \mathbf{0}$) corresponding to an identity transformation (no veiling glare).
For the target domain, the proposed LOTGMP predicts the latent maps from a noisy latent $z_t$, a target degraded latent $z_{de}^\mathcal{T}$, and the timestep $t$.
The predicted maps are injected via VGIM (Fig.~\ref{fig:veilgen}(b)), mirroring the forward degradation model (Eq.~\ref{eq:Img Forward}).
The overall training objective is a weighted combination of losses from both domains, each conditioned on a domain-specific text prompt:
\begin{equation}
\mathcal{L}_{gen} = p~\mathcal{L}_\mathcal{S} + (1-p)~\mathcal{L}_\mathcal{T}.
\label{eq:gen_loss_main}
\end{equation}

The source loss, $\mathcal{L}_\mathcal{S} = \mathbb{E} [ \|\epsilon - \epsilon_\theta(z^\mathcal{S}_t, c^\mathcal{S}, c_{txt}^\mathcal{S}, c_{vg}, t) \|_2^2 ]$, is conditioned on both the clean image $c^s$ and a text prompt $c_{txt}^\mathcal{S}$ describing only aberrations.
The target loss, $\mathcal{L}_\mathcal{T} = \mathbb{E} [ \|\epsilon - \epsilon_\theta(z^\mathcal{T}_t, \varnothing, c_{txt}^\mathcal{T}, c_{vg}, t) \|_2^2 ]$, is guided by both a text prompt $c_{txt}^\mathcal{T}$ describing the full compound degradations and the latent maps $c_{vg}$. 
The expectation $\mathbb{E}$ is taken over all relevant variables for each domain, and $p \in [0, 1]$ is a balancing hyperparameter.
\noindent\textbf{Degradation-Clean Paired Data Generation.} 
Once trained, VeilGen is employed to generate the paired dataset for training our restoration network in an offline process (Algorithm~\ref{alg:data generation}).
We use the standard notation from DDPM~\cite{ho2020denoising}, where $\beta_t \in (0, 1)$ is the variance schedule, $\alpha_t = 1 - \beta_t$, and $\bar{\alpha}_t = \prod_{i=1}^t \alpha_i$ is the cumulative product. 
To generate each degraded image $I_{de}$, we first extract latent maps $c_{vg}$ from an unpaired target-domain image $I_{de}^\mathcal{T}$ via the frozen LOTGMP.
These maps then guide a full diffusion sampling process that applies the degradation to a clean image $I_{c}$. 
This sampling process incorporates the blended strategy from~\cite{wang2025learning}, mixing conditional and unconditional predictions to enhance realism and diversity. 
This process produces a high-fidelity paired dataset in which each degraded image contains realistic compound degradations, including aberration and veiling glare.
\begin{algorithm}[tb]
  \caption{Degradation-Clean Paired Data Generation} 
  \label{alg:data generation}
  \begin{algorithmic}[1]
    \Require Denoising model $\epsilon_\theta$, VAE encoder~$\mathcal{E}$ and decoder $\mathcal{D}$, LOTGMP $\mathcal{P}$, $I_{c}$: a clean image, $I_{de}^\mathcal{T}$: a target-domain degraded image, $w$: mixture coefficient
    \State $\mathbf{z}_{de}^\mathcal{T} \gets \mathcal{E}(I_{de}^\mathcal{T})$  
    
    $\!\!\triangleright$ Set a timestep $t^{*}$ in $\mathcal{P}$
    \State $\boldsymbol{\epsilon}_{t^{*}} \sim \mathcal{N}(\mathbf{0}, \mathbf{I})$
    
    $\!\!\triangleright$ Add noise to $\mathbf{z}_{de}^\mathcal{T}$
    \State $\mathbf{z}_{t^{*}} \gets \sqrt{\bar{\alpha}_{t^{*}}} \mathbf{z}_{de}^\mathcal{T} + \sqrt{1 - \bar{\alpha}_{t^{*}}} \boldsymbol{\epsilon}_{t^{*}}$
    \State $c_{vg} \gets \mathcal{P}(\mathbf{z}_{t^{*}}, \mathbf{z}_{de}^\mathcal{T}, t^{*})$ 
    \State $\mathbf{z}_{T} \sim \mathcal{N}(\mathbf{0}, \mathbf{I})$
    \For{$t=T, \dotsc, 1$}

        $\!\!\triangleright$ Blended noise prediction
      \State $\hat{\epsilon}$ = $w$\;$\epsilon_\theta$($\mathbf{z}_{t}, \mathcal{E}(I_{c}),\varnothing, t) + (1-w)\;\epsilon_{\theta}(\mathbf{z}_{t}, \varnothing, c_{vg},t)$ 
      
      $\!\!\triangleright$ Sampling step
      \State $\epsilon \sim \mathcal{N}(\mathbf{0}, \mathbf{I})$ 
      \State $\mathbf{z}_{t-1} \gets \frac{1}{\sqrt{\alpha_t}} \left( \mathbf{z}_t-\frac{1-\alpha_t}{\sqrt{1-\bar{\alpha}_t}}\hat{\epsilon} \right) + \sqrt{\frac{1-\bar{\alpha}_{t-1}}{1-\bar{\alpha}_t}(1-\alpha_t)} \epsilon$       
    \EndFor
    \State \textbf{return} $\mathcal{D}(\mathbf{z}_{0})$

  \end{algorithmic}

\end{algorithm}
\vspace{-0.5em}
\subsection{Stage \Rmnum{2}: Forward Model Distillation}
\label{sec:stage_2}
Our objective is to use the forward degradation from VeilGen as supervision for the restoration network in Stage \Rmnum{3}.
However, invoking VeilGen during training is computationally prohibitive due to its multi-step diffusion sampling.
We therefore distill its behavior into a lightweight Distilled Degradation Network (DDN), which serves as an efficient forward model (Fig.~\ref{fig:deveiler}(a)).
Given a clean image $I_{c}$ and latent maps $c_{vg}$, the DDN is trained to regress the degraded output $I_{de}$ synthesized by VeilGen:
\begin{equation}
\mathcal{L}_{distill} = \left\| \text{DDN}(I_{c}, c_{vg}) - \text{VeilGen}(I_{c}, c_{vg}) \right\|_1.
\label{eq:distill_loss}
\end{equation}

After training, this frozen network provides an efficient forward degradation model for imposing the consistency constraint in Stage~\Rmnum{3}.
\subsection{Stage \Rmnum{3}: Reversible Restoration}
\label{sec:stage_3}
To tackle the spatial-varying veiling glare that hinders blind restoration, we develop DeVeiler, a reversible restoration network that predicts latent maps to invert the underlying degradation.
The central component is the Veiling Glare Encoder (VG-Enc), which predicts latent maps $\hat{c}_{vg} = \text{VG-Enc}(I_{de})$ representing the spatial distribution of glare-related parameters.
Simply using the predicted maps $\hat{c}_{vg}$ in a one-direction manner is ineffective due to a fundamental domain mismatch.
Latent maps optimized for conditioning iterative diffusion sampling do not transfer well to guiding the restoration process (see in Tab.~\ref{tab:latent_vg_prior}).
To bridge this gap, we introduce a reversibility-guided training paradigm that provides indirect but powerful supervision by leveraging the frozen DDN as a forward degradation model to enforce cycle consistency (Fig.~\ref{fig:deveiler}(b)).
This constraint drives $\hat{c}_{vg}$ to acquire meaningful physical interpretability, as it must parameterize two mutually inverse operations.
Within the frozen DDN, the VGIM module uses $\hat{c}_{vg}$ to impose forward modulation, applying the glare degradation to clean features.
Symmetrically, in DeVeiler, the VGCM applies an inverse modulation using the same latent maps $\hat{c}_{vg}$ to remove the glare components from degraded features. 
This structural coupling of VGIM and VGCM encourages the restoration to be a consistent inverse of the learned forward degradation, rather than a loosely correlated statistical mapping.

\noindent\textbf{Optimization Objective.} 
DeVeiler is trained end-to-end with a composite objective that balances image reconstruction fidelity and reversibility consistency:
\begin{equation}
\label{eq:total_loss}
\mathcal{L}_{total} = \mathcal{L}_{rec} + \lambda_{\text{rev}} \mathcal{L}_{rev}.
\end{equation}

Here, $\mathcal{L}_{rec}$ combines pixel-wise L1 and perceptual losses to ensure high visual fidelity between the restored and ground-truth (GT) images.
The reversibility loss $\mathcal{L}_{\text{rev}}$ enforces reversibility by constraining the estimated latent maps $\hat{c}_{vg}$ to reproduce the observed degradation through the frozen DDN:
\begin{equation}
\label{eq:phys_loss}
\mathcal{L}_{rev} = \| \text{DDN}(I_{c}, \hat{c}_{vg}) - I_{de} \|_1.
\end{equation}

\begin{table*}[t]
\footnotesize
\centering
\setlength{\tabcolsep}{0.6mm}  
\begin{tabular}{lccccccc}  %
\toprule
\multirow{2}{*}{Methods} &  
\multicolumn{3}{c}{\texttt{Screen}-SL} & 
\multicolumn{3}{c}{\texttt{Screen}-MRL} & \multicolumn{1}{c}{Overhead} \\  %
\cmidrule(lr){2-4}\cmidrule(lr){5-7}\cmidrule(lr){8-8}  
& PSNR$\uparrow$ & SSIM$\uparrow$ & LPIPS$\downarrow$ & PSNR$\uparrow$ & SSIM$\uparrow$ & LPIPS$\downarrow$ & Latency(s)$\downarrow$ \\  %
\midrule
\textit{Single Degradation Pipeline} \\  
\cellcolor{gray!15}Fast two-step\pub{ECCV22}~\cite{eboli2022fast}       & \cellcolor{gray!15}16.62 (73.5\%) & \cellcolor{gray!15}0.539 (35.3\%) & \cellcolor{gray!15}0.674 (61.3\%) & \cellcolor{gray!15}14.94 (78.3\%) & \cellcolor{gray!15}0.610 (22.3\%) & \cellcolor{gray!15}0.504 (40.3\%) & \cellcolor{gray!15}0.390 \\  
SwinIR\pub{ICCVW21}~\cite{liang2021swinir}       & 18.18 (62.0\%)     & 0.686 (6.27\%)     & 0.298 (12.4\%)     & 19.34 (40.2\%)     & 0.722 (3.32\%)     & 0.354 (15.0\%)     & 0.782 \\  
\cellcolor{gray!15}NAFNET\pub{ECCV22}~\cite{chen2022simple} & \cellcolor{gray!15}18.75 (56.7\%) & \cellcolor{gray!15}0.684 (6.58\%) & \cellcolor{gray!15}0.363 (28.1\%) & \cellcolor{gray!15}18.91 (45.8\%) & \cellcolor{gray!15}0.723 (3.18\%) & \cellcolor{gray!15}0.377 (20.2\%) & \cellcolor{gray!15}0.353 \\  
DiffBIR\pub{ECCV24}~\cite{lin2024diffbir}       & 17.95 (63.9\%)     & 0.621 (17.4\%)     & 0.398 (34.4\%)     & 18.70 (48.4\%)     & 0.625 (19.4\%)     & 0.412 (26.9\%)     & 66.13 \\  
\hdashline
\textit{Cascaded Pipeline} \\  
\cellcolor{gray!15}SwinIR\pub{ICCVW21}~\cite{liang2021swinir}+RIDCP$^{*}$\pub{CVPR23}~\cite{wu2023ridcp}       & \cellcolor{gray!15}20.06 (41.4\%) & \cellcolor{gray!15}0.695 (4.89\%) & \cellcolor{gray!15}0.385 (32.2\%) & \cellcolor{gray!15}18.66 (48.8\%) & \cellcolor{gray!15}0.692 (7.80\%) & \cellcolor{gray!15}0.456 (34.0\%) & \cellcolor{gray!15}1.240 \\  
SwinIR\pub{ICCVW21}~\cite{liang2021swinir}+DiffDehaze$^{*}$\pub{CVPR25}~\cite{wang2025learning} & 19.31 (50.7\%)     & 0.642 (13.6\%)     & 0.347 (24.8\%)     & 18.28 (53.1\%)     & 0.603 (23.7\%)     & 0.392 (23.2\%)     & 92.54 \\  
\cellcolor{gray!15}SwinIR\pub{ICCVW21}~\cite{liang2021swinir}+Flare7K++$^{*}$\pub{TPAMI2024}~\cite{dai2024flare7k++}      & \cellcolor{gray!15}\blue{\underline{21.67}} (15.1\%) & \cellcolor{gray!15}\blue{\underline{0.723}} (0.86\%) & \cellcolor{gray!15}0.297 (12.1\%) & \cellcolor{gray!15}\blue{\underline{20.74}} (17.4\%) & \cellcolor{gray!15}\blue{\underline{0.745}} (0.13\%) & \cellcolor{gray!15}0.336 (10.4\%) & \cellcolor{gray!15}0.990 \\  
\hdashline
\textit{Domain Adaptation Pipeline} \\  
\cellcolor{gray!15}CycleGAN\pub{ICCV17}~\cite{CycleGAN2017} & \cellcolor{gray!15}18.20 (61.8\%) & \cellcolor{gray!15}0.558 (30.6\%) & \cellcolor{gray!15}0.649 (59.8\%) & \cellcolor{gray!15}17.80 (58.0\%) & \cellcolor{gray!15}0.623 (19.7\%) & \cellcolor{gray!15}0.485 (37.9\%) & \cellcolor{gray!15}0.208 \\  
UCL-Dehaze\pub{TIP24}~\cite{wang2024ucl}       & 18.55 (58.6\%)     & 0.581 (25.5\%)      & 0.554 (52.9\%)     & 18.69 (48.5\%)     & 0.688 (8.43\%)     & 0.451 (33.3\%)      & 0.208 \\  
\cellcolor{gray!15}DiffDehaze\pub{CVPR25}~\cite{wang2025learning} & \cellcolor{gray!15}18.87 (55.4\%) & \cellcolor{gray!15}0.657 (11.0\%) & \cellcolor{gray!15}0.435 (40.0\%) & \cellcolor{gray!15}18.55 (50.1\%) & \cellcolor{gray!15}0.645 (15.7\%) & \cellcolor{gray!15}0.426 (29.3\%) & \cellcolor{gray!15}91.76 \\  
QDMR\pub{OLT2025}~\cite{jiang2025qdmr}       & 18.45 (59.5\%)     & 0.681 (7.05\%)     & \blue{\underline{0.291}} (10.3\%)     & 20.67 (18.7\%)     & 0.725 (2.90\%)     & \blue{\underline{0.315}} (4.44\%)     & 0.428 \\  
\cellcolor{gray!15}\textbf{DeVeiler (Ours)} & \cellcolor{gray!15}\red{\textbf{22.38}} (0.00\%)     & \cellcolor{gray!15}\red{\textbf{0.729}} (0.00\%)     & \cellcolor{gray!15}\red{\textbf{0.261}} (0.00\%)     & \cellcolor{gray!15}\red{\textbf{21.57}} (0.00\%)     & \cellcolor{gray!15}\red{\textbf{0.746}} (0.00\%)     & \cellcolor{gray!15}\red{\textbf{0.301}} (0.00\%)     & \cellcolor{gray!15}0.387 \\  
\bottomrule
\end{tabular}
\vskip-2.5ex
\caption{\textbf{Quantitative comparison on the \texttt{Screen-Compound} domain (SL and MRL systems).} $^{*}$ denotes using the pretrained model. Latency(s) is computed on images with the size of $1280{\times}1920$ with an NVIDIA A100 GPU. \red{\textbf{Red}} and \blue{\underline{blue}} indicate the best and the second best performance, respectively. The numbers in parentheses denote the improvement rate of our method compared to others.}
\vspace{-4.0mm}
\label{tab:results_screen}
\end{table*}

\noindent\textbf{Two-Phase Training.} To effectively address the domain gap, we propose a two-phase training strategy.
In Phase~\Rmnum{1} (pre-training), DeVeiler is first trained on the source domain to establish a solid baseline for aberration correction.
In Phase~\Rmnum{2} (fine-tuning), it is further adapted using a hybrid dataset that mixes synthetic paired data generated by VeilGen with the original source-domain pairs.
This hybrid approach is critical, acting as a powerful regularizer: it allows the network to learn the new compound degradation from the synthetic pairs, while the source domain data preserves the foundational aberration correction and prevents overfitting to the synthetic-only distribution.

\section{Experiments and Results}
\label{sec:exp}
\subsection{Experiment Settings}
\label{subsec:exp_set}
\noindent\textbf{Datasets.} 
We verify our method on two distinct optical systems: a large-aperture single lens (SL) and a metasurface-refractive hybrid lens (MRL). 
We first constructed a source domain of paired, aberration-only data. 
Lacking the precise optical parameters required for simulation, we follow prior work~\cite{chen2024hybrid,zhang2025degradation} and adopt a screen-capture setup to re-capture clean images from the DIV2K dataset~\cite{timofte2017ntire}, yielding $170$ (SL) and $125$ (MRL) pairs.
We then collected two distinct Target Domains for adaptation, adhering to a strict ``Source + one Target" setup.
The \texttt{Screen-Compound} domain used the same screen-capture method but with artificial lighting to induce compound degradation, providing GT pairs for $50$ training and $42{/}25$ (SL/MRL) test images.
The \texttt{Realworld-Compound} domain consists of real-world captures exhibiting natural compound degradation with no GT, providing $50$ training and $51{/}11$ (SL/MRL) test images.
Finally, we use $500$ clean images from the Flickr2K dataset~\cite{timofte2017ntire} to generate $500$ corresponding target-domain degraded pairs.
All data across both systems are captured at a resolution of $1280{\times}1920$.
See the supplement for further data construction details.
%
\setlength{\tabcolsep}{0.8mm}{  
\begin{table}[t]\footnotesize
\centering
\begin{tabular}{lcc}  
\toprule
Methods       & \texttt{Realworld}-SL & \texttt{Realworld}-MRL \\  
\midrule
\multicolumn{3}{l}{\textit{Single Degradation Pipeline}} \\
\cellcolor{gray!15}Fast two-step~\cite{eboli2022fast}       & \cellcolor{gray!15}0.356/2.942/5.821 & \cellcolor{gray!15}0.388/2.687/5.352 \\
SwinIR~\cite{liang2021swinir}       & 0.424/3.518/5.710 & 0.374/3.191/6.696 \\
\cellcolor{gray!15}NAFNET~\cite{chen2022simple} & \cellcolor{gray!15}0.374/3.598/5.535 & \cellcolor{gray!15}0.398/3.238/5.512 \\
DiffBIR~\cite{lin2024diffbir}       & 0.404/3.476/5.770 & 0.367/2.761/6.251 \\
\hdashline
\multicolumn{3}{l}{\textit{Cascaded Pipeline}} \\
\cellcolor{gray!15}SwinIR~\cite{liang2021swinir}+RIDCP$^{*}$~\cite{wu2023ridcp}       & \cellcolor{gray!15}0.386/3.755/5.810 & \cellcolor{gray!15}0.399/3.290/6.502 \\
SwinIR~\cite{liang2021swinir}+DiffDehaze$^{*}$~\cite{wang2025learning} & \blue{\underline{0.573}}/3.679/6.141 & \blue{\underline{0.437}}/\blue{\underline{3.542}}/\blue{\underline{5.230}} \\
\cellcolor{gray!15}SwinIR~\cite{liang2021swinir}+Flare7K++$^{*}$~\cite{dai2024flare7k++}      & \cellcolor{gray!15}0.459/3.528/5.622 & \cellcolor{gray!15}0.387/3.319/6.336 \\
\hdashline
\multicolumn{3}{l}{\textit{Domain Adaptation Pipeline}} \\
\cellcolor{gray!15}CycleGAN~\cite{CycleGAN2017} & \cellcolor{gray!15}0.451/2.961/5.300 & \cellcolor{gray!15}0.418/3.319/\red{\textbf{3.836}} \\
UCL-Dehaze~\cite{wang2024ucl}       & 0.297/2.763/8.322 & 0.343/2.963/5.644 \\
\cellcolor{gray!15}DiffDehaze~\cite{wang2025learning} & \cellcolor{gray!15}0.406/\blue{\underline{3.982}}/6.476 & \cellcolor{gray!15}0.428/3.476/6.802 \\
QDMR~\cite{jiang2025qdmr}       & 0.405/3.864/\blue{\underline{4.773}} & 0.376/3.337/5.509 \\
\cellcolor{gray!15}\textbf{DeVeiler (Ours)} & \cellcolor{gray!15}\red{\textbf{0.607}/\textbf{3.987}/\textbf{4.448}} & \cellcolor{gray!15}\red{\textbf{0.440}}/\red{\textbf{3.586}}/5.296 \\
\bottomrule
\end{tabular}
\vspace{-3.5mm}
\caption{\textbf{Quantitative comparison (CLIPIQA$\uparrow$/Q-Align$\uparrow$ /NIQE$\downarrow$) on the \texttt{Realworld-Compound} domain (SL and MRL systems).} $^{*}$ denotes using the pretrained model.}
\label{tab:results_realworld}
\vspace{-5mm}
\end{table}}

\noindent\textbf{Implementation Details.} 
VeilGen builds upon pre-trained Stable Diffusion v2-1~\cite{rombach2022high} and is optimized with AdamW~\cite{Ilya2019adamw} at a learning rate of $1{\times}10^{-5}$ for 9k iterations with a batch size of 16.
The probability parameter $p$ is set to $0.3$ and the mixture coefficient $w$ to $0.85$, consistent with common practices~\cite{wang2025learning}.
$t^{*}$ is set to $0$.
DeVeiler is trained in two phases using Adam~\cite{kingma2014adam}: Phase~\Rmnum{1} pre-trains on source-domain pairs, and Phase~\Rmnum{2} fine-tunes on a hybrid dataset of $500$ synthetic pairs and source pairs. 
Each encoder and decoder consists of three groups of ResBlocks, while the bottleneck is constructed with RSTB layers from SwinIR~\cite{liang2021swinir}.
The balance weight $\lambda_{\text{rev}}$ is set to $1.0$. 
All experiments are conducted on an NVIDIA A100 GPU.
More details are delivered in the supplementary material.

\begin{figure*}[!t]
  \centering
  \includegraphics[width=0.99\linewidth]{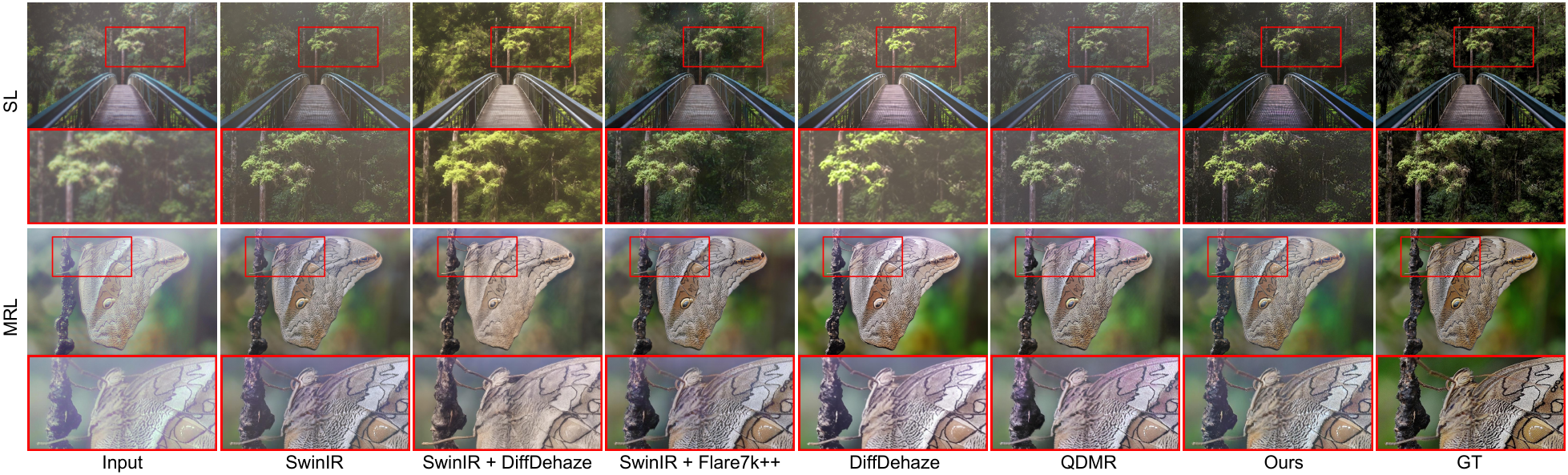}
  \vskip-2.0ex
  \caption{\textbf{Visual results on \texttt{Screen-Compound} domain.} Zoom in for the best view.}
  \label{fig:visual_screen}
  \vspace{-1.0em}
\end{figure*}

\begin{figure*}[!t]
  \centering
  \includegraphics[width=0.99\linewidth]{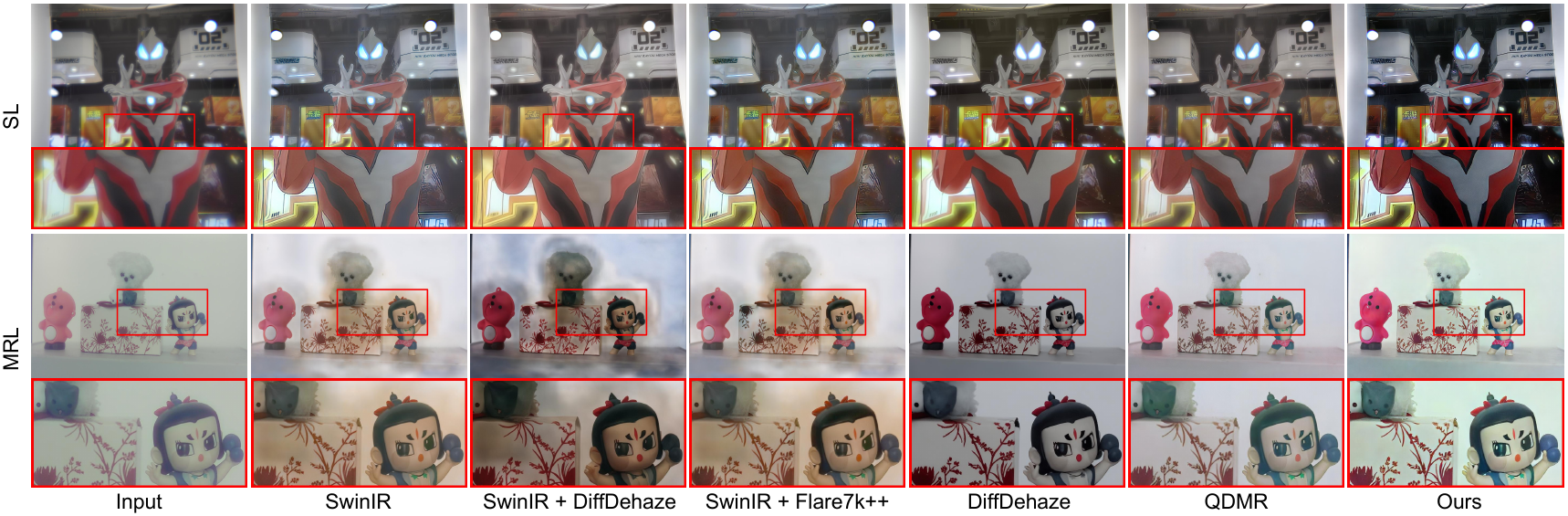}
  \vskip-2.0ex
  \caption{\textbf{Visual results on \texttt{Realworld-Compound} domain.} Zoom in for the best view.}
  \label{fig:visual_realworld}
  \vspace{-1.5em}  
\end{figure*}

\subsection{Comparison with State-of-the-Art Methods}
\label{subsec:comparison}
We compare against three pipelines in Tab.~\ref{tab:results_screen} and Tab.~\ref{tab:results_realworld}.
First, we evaluate single-degradation baselines, which include the blind CAC method Fast two-step~\cite{eboli2022fast} as well as well-known image restoration models (SwinIR~\cite{liang2021swinir}, NAFNet~\cite{chen2022simple}, DiffBIR~\cite{lin2024diffbir}) re-trained on source domain pairs.
Second, cascaded pipelines combine the re-trained SwinIR with pre-trained dehazing (RIDCP~\cite{wu2023ridcp}, DiffDehaze~\cite{wang2025learning}) or flare removal (Flare7K++~\cite{dai2024flare7k++}) models to validate the distinct nature of veiling glare.
Finally, the Domain Adaptation (DA) methods are all retrained on our source and target domains for a fair comparison. 
This includes CycleGAN~\cite{CycleGAN2017}, UCL-Dehaze~\cite{wang2024ucl}, the aberration-specific method QDMR~\cite{jiang2025qdmr}, and a re-trained DiffDehaze~\cite{wang2025learning}.
\noindent\textbf{Results on \texttt{Screen-Compound} Domain.} We first conduct a quantitative evaluation on the \texttt{Screen-Compound} domain, which provides paired GT. 
We employ full-reference metrics: PSNR and SSIM~\cite{wang2004image} to measure fidelity, and LPIPS~\cite{zhang2018unreasonable} to assess perceptual quality. 
As shown in Tab.~\ref{tab:results_screen}, our DeVeiler significantly outperforms all baselines across all metrics.
Single degradation pipeline, trained only on the source domain, fails to generalize to the unseen veiling glare. 
Cascaded pipelines also prove insufficient. 
We use the strongest single-degradation model, SwinIR, as the CAC base, followed by pre-trained dehazing or flare removal models. 
These naive cascades yield unsatisfactory results: dehazing models severely harm color fidelity, while the flare removal model shows negligible perceptual gains.
This confirms the distinct nature of veiling glare compared to these other degradations.
Furthermore, the DA pipeline produces sub-optimal results due to its physically-agnostic designs.
In contrast, DeVeiler leverages the latent veiling glare maps ($c_{vg}$) via the mutually inverse VGIM/VGCM structure, enabling it to precisely reverse the degradation.

Visual comparisons in Fig.~\ref{fig:visual_screen} on both SL and MRL systems confirm these quantitative findings. 
For example, in the first row, the results of DeVeiler consistently restore high color fidelity, and in the second row, they successfully remove severe veiling glare while preserving fine details. In contrast, competing methods exhibit clear color shifts, residual glare, or loss of detail.

\noindent\textbf{Results on \texttt{Realworld-Compound} Domain.} We further evaluate the performance of DeVeiler on the challenging \texttt{Realworld-Compound} domain using both SL and MRL systems.
As no GT is available, we employ three widely-used no-reference IQA metrics: CLIPIQA~\cite{wang2023exploring}, Q-Align~\cite{wu2023q}, and NIQE~\cite{mittal2012making}.
Tab.~\ref{tab:results_realworld} shows that DeVeiler achieves excellent performance across the majority of metrics, verifying its generalizability in handling complex, real-world degradations.
In addition, we provide visual comparisons in Fig.~\ref{fig:visual_realworld}. 
DeVeiler consistently restores fine details and color fidelity while introducing minimal artifacts, whereas other methods struggle with residual glare and artifacts.
Notably, the first row example highlights that aberration and veiling glare are a more pervasive challenge in compact systems than severe flare or ghosting, which are situational.
These experimental results demonstrate the robustness of DeVeiler in realistic scenarios. We refer the reader to the supplementary material for more visual results.

\subsection{Ablation Studies}
To analyze the effectiveness of each component, we conduct ablation studies on the \texttt{Screen-Compound} SL dataset, maintaining consistent settings for fair comparison.

\label{subsec:Ablation Studies}
\noindent\textbf{Ablation Study for VeilGen.} We evaluate the impact of the LOTGMP module by training DeVeiler (\textit{w/o} VGCM) on data generated from different VeilGen variants.
As shown in Tab.~\ref{tab:ablation_veilgen}, removing LOTGMP or discarding its SD-based prior input ($z_t$) consistently degrades performance, indicating that the SD-guided latent input is essential for adaptive prior learning ($c_{vg}$) and realistic data synthesis.

\begin{table}[t]
\small
\centering
\setlength{\tabcolsep}{5mm}
\resizebox{0.48\textwidth}{!}{  
  \begin{tabular}{ccccc}  
    \toprule
    LOTGMP & SD-based prior  & PSNR$\uparrow$ & SSIM$\uparrow$ & LPIPS$\downarrow$ \\  
    \cmidrule(lr){1-5}  
    \ding{55} & \ding{55} & 20.82  & 0.708 & 0.273 \\  
    \cellcolor{gray!15}\checkmark & \cellcolor{gray!15}\ding{55} & \cellcolor{gray!15}21.39 & \cellcolor{gray!15}0.708 & \cellcolor{gray!15}0.268 \\  
    \checkmark & \checkmark & \textbf{\red{21.56}} & \textbf{\red{0.712}} & \textbf{\red{0.264}} \\  
    \bottomrule
  \end{tabular}
}
\vspace{-3.0mm}
\caption{Ablation study on the proposed VeilGen.}
\label{tab:ablation_veilgen}
\vspace{-3.0mm}
\end{table}
\begin{table}[t]
\small
\centering
\setlength{\tabcolsep}{4.5mm}
\resizebox{0.48\textwidth}{!}{  
  \begin{tabular}{lccc}  
    \toprule
    Methods & PSNR$\uparrow$ & SSIM$\uparrow$ & LPIPS$\downarrow$ \\  
    \midrule
    CycleGAN        & 18.85 & 0.706 & 0.326 \\  
    \cellcolor{gray!15}Degradation Transfer    & \cellcolor{gray!15}17.64 & \cellcolor{gray!15}0.686 & \cellcolor{gray!15}0.353 \\  
    HazeGen     & 20.82 & 0.709 & 0.273 \\  
    \cellcolor{gray!15}\textbf{VeilGen (Ours)}     & \cellcolor{gray!15}\textbf{\red{21.56}} & \cellcolor{gray!15}\textbf{\red{0.712}} & \cellcolor{gray!15}\textbf{\red{0.264}} \\  
    \bottomrule
  \end{tabular}
}
\vspace{-3.0mm}
\caption{Ablation study on data generation.}
\label{tab:ablation_data_generation}
\vspace{-3.0mm}
\end{table}
\begin{figure}[!t]
  \centering
  \includegraphics[width=0.99\linewidth]{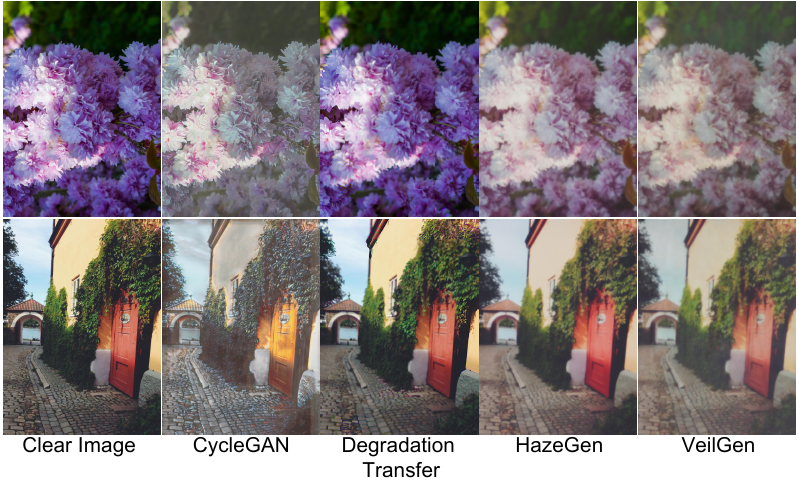}
  \vskip-2.0ex
  \caption{Visual comparison of data generation methods.}
  \label{fig:ablation_data_gen_vis}
  \vspace{-1.0em}
\end{figure}

\noindent\textbf{Effectiveness of VeilGen.}
We compare VeilGen with competitive data generation methods (CycleGAN~\cite{CycleGAN2017}, DegradationTransfer~\cite{chen2021extreme_quality}, HazeGen~\cite{wang2025learning}). 
For a fair comparison, an identical restoration network (DeVeiler (\textit{w/o} VGCM)) is trained on the data generated by each method. 
As illustrated in Tab.~\ref{tab:ablation_data_generation}, training the restoration network on data from VeilGen yields superior performance, indicating that its generated pairs more accurately capture the target degradation characteristics.
The visual comparison (Fig.~\ref{fig:ablation_data_gen_vis}) further shows that VeilGen synthesizes realistic compound degradations while other methods fail.

\begin{figure}[!t]
  \centering
  \includegraphics[width=0.99\linewidth]{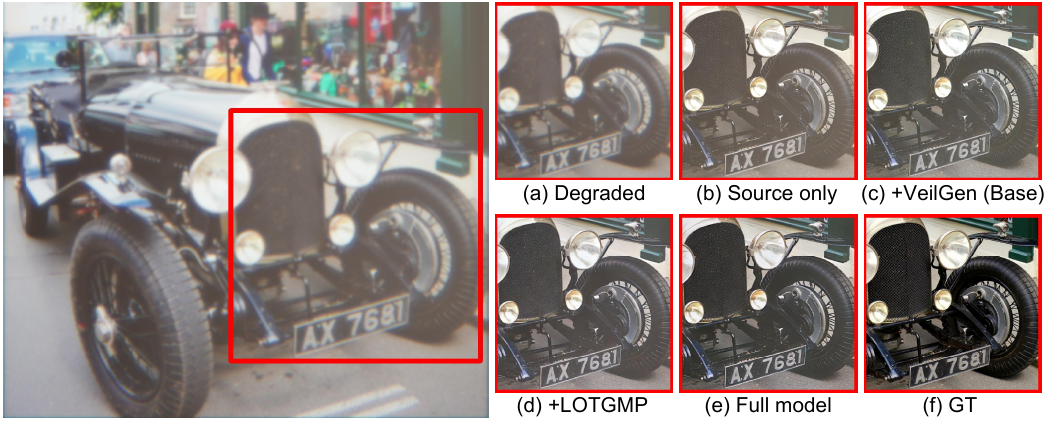}
  \vskip-2.0ex
  \caption{Visual ablation of core components.}
  \label{fig:ablation_vis}
  \vspace{-1.0em}
\end{figure}

\noindent\textbf{Ablation Study for DeVeiler.} We next ablate the method for using the latent maps ($c_{vg}$) in DeVeiler (Tab.~\ref{tab:latent_vg_prior}). 
The results show that naive unidirectional injection (via Concat or VGCM) provides no benefit. 
We attribute this failure to a fundamental domain mismatch: diffusion-optimized latent maps are ill-suited for conventional restoration networks. 
This mismatch is resolved by our reversible paradigm, where the bidirectional VGIM/VGCM structure enforces pathway consistency and unlocks performance gains.
\noindent\textbf{Visual Ablation of Core Components.} Fig.~\ref{fig:ablation_vis} shows the cumulative effect of our core components. 
The source-only baseline fails to remove veiling glare. 
Adding LOTGMP improves data generation (consistent with Tab.~\ref{tab:ablation_veilgen}), and activating the full bidirectional DeVeiler further eliminates residual artifacts, producing results closest to the GT.

\begin{table}[t]
\small
\centering
\setlength{\tabcolsep}{4mm}
\resizebox{0.48\textwidth}{!}{  
  \begin{tabular}{lccc}  
    \toprule
    Methods & PSNR$\uparrow$ & SSIM$\uparrow$ & LPIPS$\downarrow$ \\  
    \midrule
    Baseline (\textit{w/o} Latent maps)                    & 21.56 & 0.712 & 0.264 \\  %
    \cellcolor{gray!15}Concat (Unidirectional)                & \cellcolor{gray!15}21.12 & \cellcolor{gray!15}0.671 & \cellcolor{gray!15}0.317 \\  
    VGCM (Unidirectional)   & 20.83 & 0.712 & 0.264 \\  
    \cellcolor{gray!15}Concat (Bidirectional)   & \cellcolor{gray!15}20.65 & \cellcolor{gray!15}0.662 & \cellcolor{gray!15}0.329 \\  
    \textbf{VGIM/VGCM (Bidirectional)} & \textbf{\red{22.38}}  & \textbf{\red{0.729}} & \textbf{\red{0.261}} \\  
    \bottomrule
  \end{tabular}
}
\vspace{-3mm}
\caption{Ablation study on the LOTGMP usage.}
\label{tab:latent_vg_prior}
\vspace{-2.0mm}
\end{table}

\begin{figure}[!t]
  \centering
  \includegraphics[width=0.99\linewidth]{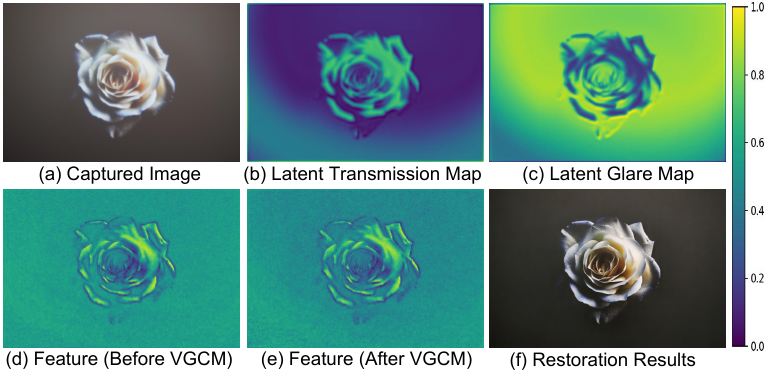}
  \vskip-1.0ex
  \caption{Interpretability of the reversible mechanism.}
  \label{fig:Interpretability}
  \vspace{-1.0em}
\end{figure}

\noindent\textbf{Analysis of the VGIM/VGCM Mechanism.}
Fig.~\ref{fig:Interpretability}(b)-(e) visualizes internal feature maps of DeVeiler. 
The network accurately estimates latent components, identifying low-transmittance and high-glare regions in the input. 
After modulation by the VGCM, activations in the glare-affected regions are effectively suppressed. 
This verifies that DeVeiler learns an interpretable restoration process, where the VGCM module specifically targets and removes the glare component, rather than performing a black-box mapping.
\section{Conclusion}
\label{sec:Conclusion}
In this paper, we address the problem of compound degradation involving co-existing aberration and veiling glare in simplified optical systems. 
Our physics-informed framework unifies data generation (VeilGen) and restoration (DeVeiler) through a shared latent map realized via bidirectional modulation. 
Experimental results show that our method performs favorably compared to other methods.
We hope our work can inspire the development of lightweight, high-quality imaging systems.

\clearpage
\section*{Acknowledgment}

This research was funded by the Natural Science Foundation of Zhejiang Province (Grant No. LZ24F050003), National Natural Science Foundation of China (Grant Nos. 12174341 and 62473139), Hunan Provincial Research and Development Project (Grant 2025QK3019), and State Key Laboratory of Autonomous Intelligent Unmanned Systems (the opening project number ZZKF2025-2-10).
{
    \small
    \bibliographystyle{ieeenat_fullname}
    \bibliography{main}
}

The supplementary material is organized as follows. 
First, \S\ref{sec:data_setup} details the procedures for data acquisition and usage across different domains.
Next, \S\ref{sec:supp_motivation} presents a detailed motivation analysis regarding the learning of latent veiling glare maps. 
\S\ref{sec:Implementation Details} provides further implementation details, including training configurations and hyperparameters. 
In \S\ref{sec:discussion}, we discuss the broader societal impacts of our work, along with its limitations and potential future directions. 
Finally, \S\ref{sec:More Results} demonstrates the high data efficiency of our domain adaptation framework and presents additional visual comparisons.

\section{Detailed Data Collection and Usage}
\label{sec:data_setup}
This section details the three-domain dataset structure.
We evaluate our method on two distinct optical systems: the large-aperture single lens (SL) and the metasurface-refractive hybrid lens (MRL), as shown in Fig.~\ref{fig:exp_setup}(a)-(b) and Fig.~\ref{fig:rebuttal optical infor}. 
A summary of acquisition setups, illumination conditions, and dataset splits is provided in Tab.~\ref{tab:dataset_summary}.
To illustrate the degradation components, Fig.~\ref{fig:cap_data} shows a comparison among the clear image, the aberration-only capture (Source domain), and the compound case with additional veiling glare (\texttt{Sreen-Compound} domain). 
This reveals that veiling glare degrades the image through global contrast reduction and color shifts, compounding the intrinsic optical blur.
While identical scenes are presented here for direct comparison, we ensure strictly disjoint scene contents across all dataset splits in our experiments to prevent data leakage.
All procedures described below are conducted independently for each system, resulting in two parallel datasets.

\subsection{Data Acquisition Setup}
We employ two distinct data acquisition setups: a controlled monitor-based setup for the Source and \texttt{Screen-Compound} domains, and an in-the-wild capture setup for the \texttt{Realworld-Compound} domain.

\begin{figure*}[!t]
  \centering
  \includegraphics[width=0.99\linewidth]{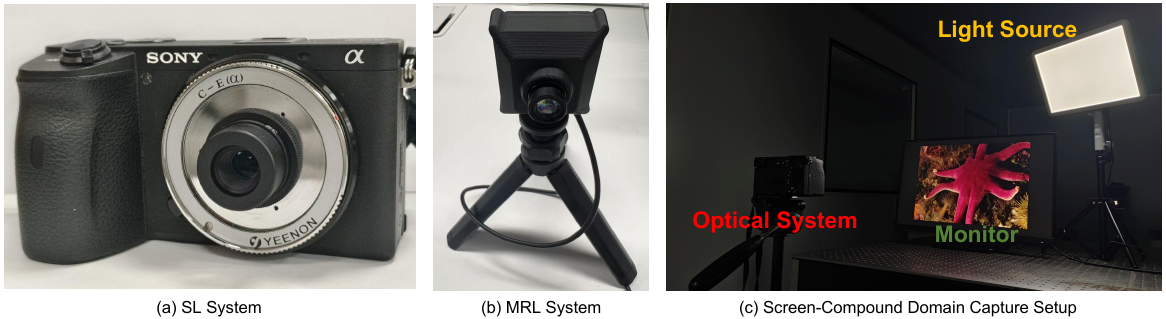}
  \caption{(a) A large-aperture single-lens (SL) system. (b) A metasurface–refractive hybrid-lens (MRL) system. (c) The capture setup for the \texttt{Screen-Compound} domain.}
  \label{fig:exp_setup}
\end{figure*}

\begin{figure*}[!t]
    \centering

    \includegraphics[width=0.99\linewidth]{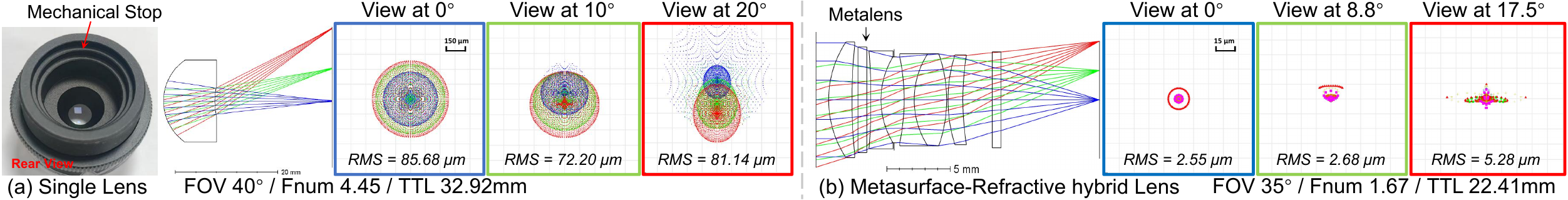}

    \caption{\textbf{Optical specifications and simulated performance.} Simulated ray tracing layouts and corresponding RMS spot diagrams for (a) the SL imaging system and (b) the MRL imaging system.}
    \label{fig:rebuttal optical infor}
\end{figure*}

\begin{figure*}[!t]
  \centering
  \includegraphics[width=0.58\linewidth]{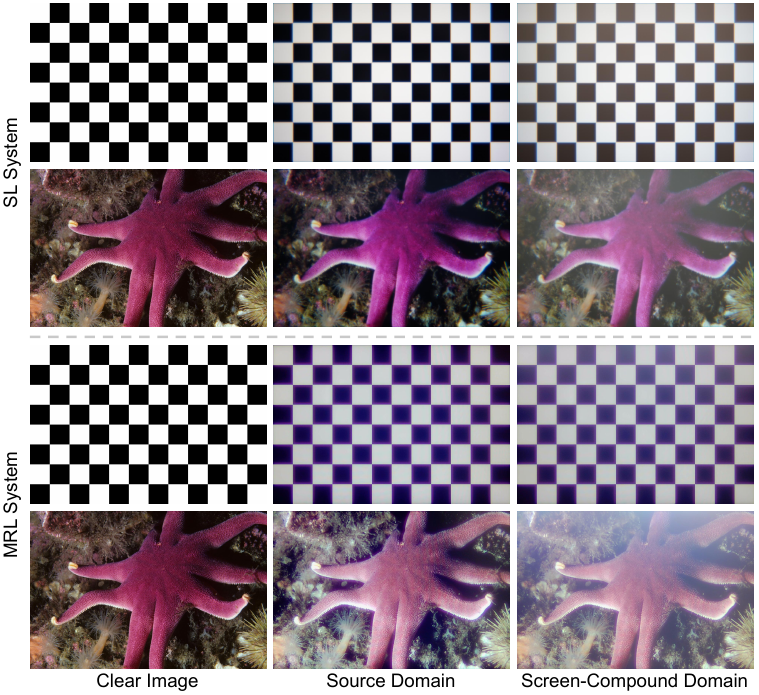}
  \caption{Visual Decomposition of Compound Degradation.}
  \label{fig:cap_data}
\end{figure*}

\noindent\textbf{Monitor-based Setup.} 
We use a screen-capture setup~\cite{chen2024hybrid,zhang2025degradation} for both the Source and \texttt{Screen-Compound} domains.
This approach is chosen because precise lens parameters for accurate simulation are not available.
In this setup, a high-resolution monitor displays the GT images, and the optical system is placed on a controlled mounting setup to capture the screen.
The illumination conditions are carefully controlled to distinguish the two domains:
\begin{itemize}
    \item \textbf{Source Domain:} The environment is kept completely dark to avoid any ambient light, ensuring that the captured images contain only intrinsic optical aberrations without veiling glare.
    \item \textbf{\texttt{Screen-Compound} Domain:} Under the same geometric alignment, we introduce an external light source (Fig.~\ref{fig:exp_setup}(c)) to generate compound degradation.
    This source is placed off the optical axis and remains outside the field of view. 
    This configuration simulates the common real-world scenario where invisible external light induces diffuse veiling glare.
    We employ this standardized lighting setup to establish a controlled benchmark.
    Despite lacking the variability of natural environments, this setup provides the essential paired data required for full-reference validation.
    In contrast, the \texttt{Realworld-Compound} domain (detailed below) encompasses random, diverse lighting conditions to verify the method's generalization capability.

\end{itemize}

\begin{table*}[t]
\centering
\renewcommand{\arraystretch}{1.25} 
\setlength{\tabcolsep}{7pt} 
\resizebox{\textwidth}{!}{%
\begin{tabular}{lcccccc}
\toprule
\rowcolor{white} 
\multirow{2}{*}{\textbf{Domain}} & \textbf{Acquisition} & \textbf{Lighting} & \textbf{GT} & \multicolumn{2}{c}{\textbf{Data Quantity (SL / MRL)}} & \multirow{2}{*}{\textbf{Usage Setting}} \\
\cmidrule(lr){5-6}
\rowcolor{white}
 & \textbf{Setup} & \textbf{Condition} & \textbf{Avail.} & \textbf{Training} & \textbf{Testing} & \\
\midrule

\rowcolor{Gray}
\textbf{Source} & Monitor-based & Dark  & \cmark & 170 / 125 & --- & \textbf{Train:} Supervised (Aberration Only) \\

\addlinespace[0.2em] 

\textbf{\texttt{Screen-Compound}} & Monitor-based & Artificial light & \cmark\textsuperscript{\textdagger} & 50 / 50 & 42 / 25 & \begin{tabular}[c]{@{}l@{}}\textbf{Train:} Unpaired Adaptation (GT Discarded)\\ \textbf{Test:} Full-Ref. Metrics \& Visual Comp.\end{tabular} \\

\addlinespace[0.2em]

\rowcolor{Gray}
\textbf{\texttt{Realworld-Compound}} & In-the-wild & Uncontrolled & \xmark & 50 / 50 & 51 / 11 & \begin{tabular}[c]{@{}l@{}}\textbf{Train:} Unpaired Adaptation\\ \textbf{Test:} No-Ref. Metrics \& Visual Comp.\end{tabular} \\

\bottomrule
\multicolumn{7}{l}{\footnotesize \textsuperscript{\textdagger} GTs are available but explicitly discarded during training to ensure strictly unpaired adaptation.}
\end{tabular}%
}
\caption{\textbf{Summary of Datasets and Usage Setting.} We evaluate our method on two independent optical systems: Large-aperture Single Lens (SL) and Metasurface-Refractive Lens (MRL). The table details the acquisition setup, lighting conditions, data splits, and specific usage for each domain.}
\label{tab:dataset_summary}
\end{table*}

\noindent\textbf{In-the-wild Capture Setup.} 
We collect diverse real-world scenes in uncontrolled environments to form the \texttt{Realworld-Compound} domain.
These images are captured under naturally varying illumination, where the intensity and distribution of light are not regulated. 
As a result, the collected data reflect the actual imaging behavior of the optical system in everyday scenarios in which glare arises naturally.
\paragraph{Geometric Alignment.}
To ensure precise pixel-level correspondence for paired data, a two-stage geometric correction process is applied independently for each system.
\begin{itemize}
    \item \textbf{Lens Distortion Correction:} Intrinsic parameters are calibrated using the MATLAB Camera Calibrator~\cite{MATLAB}. Chessboard images captured from multiple viewpoints allow computation of a distortion map that removes nonlinear distortions such as barrel or pincushion effects.
    \item \textbf{Affine Alignment:} Residual planar misalignment between the sensor and the monitor is corrected using an affine transformation. We extract $N$ corner correspondences $\{(x_i, y_i) \to (x'_i, y'_i)\}$ from a high-definition chessboard displayed on the monitor. The transformation is modeled directly in homogeneous coordinates as:
    $$
    \begin{bmatrix} x'_i \\ y'_i \\ 1 \end{bmatrix} = \mathbf{M}_{\text{aff}} \begin{bmatrix} x_i \\ y_i \\ 1 \end{bmatrix} = \begin{bmatrix} m_{11} & m_{12} & t_x \\ m_{21} & m_{22} & t_y \\ 0 & 0 & 1 \end{bmatrix} \begin{bmatrix} x_i \\ y_i \\ 1 \end{bmatrix}
    $$
    The affine parameters are estimated by minimizing the sum of squared reprojection errors:
    $$
    \min_{m_{ij}, t} \sum_{i=1}^{N} \left\| \begin{bmatrix} m_{11} & m_{12} \\ m_{21} & m_{22} \end{bmatrix} \begin{bmatrix} x_i \\ y_i \end{bmatrix} + \begin{bmatrix} t_x \\ t_y \end{bmatrix} - \begin{bmatrix} x'_i \\ y'_i \end{bmatrix} \right\|^2
    $$
\end{itemize}

This correction is strictly applied to the Source and \texttt{Screen-Compound} domains to ensure high-fidelity alignment for quantitative metrics. Conversely, for the \texttt{Realworld-Compound} domain, geometric correction is intentionally omitted.
By operating directly on uncorrected RGB images containing native distortion, we explicitly validate the robustness of the model in practical, calibration-free deployment scenarios.

\subsection{Domain Definitions and Usage}

We define three domains. The Source domain establishes baseline capabilities for aberration synthesis and correction. For unpaired adaptation, each target domain (\texttt{Screen-Compound} and \texttt{Realworld-Compound}) is treated separately, ensuring the model adapts to each domain independently.

\paragraph{Source Domain (Paired, Aberration-only).}
\begin{itemize}
    \item \textbf{Acquisition:} Using the aligned screen-capture setup, clean images from the DIV2K dataset~\cite{timofte2017ntire} are re-captured in a dark environment. This ensures that only the intrinsic spatial-varying aberrations are recorded.
    \item \textbf{Data Split:} This domain yields $170$ pairs for the SL system and $125$ pairs for the MRL system.
    \item \textbf{Usage:} These pairs provide fully supervised supervision for modeling and correcting spatial-varying aberrations.
\end{itemize}

\paragraph{Target Domain: \texttt{Screen-Compound} (Unpaired Training, Paired Testing).}
This domain introduces controlled veiling glare and serves a dual purpose for training and evaluation.
\begin{itemize}
    \item \textbf{Acquisition:} The process follows the Source domain setup, except that a single external light source is directed toward the lens (Fig.~\ref{fig:exp_setup}), inducing compound degradation from both aberrations and veiling glare.
    \item \textbf{Data Split:} We collect $50$ training images and $42$ test images for the SL system ($50$ training, $25$ test for MRL). All images are captured with corresponding GT pairs.
    \item \textbf{Usage:} The dataset is strictly split:
        \begin{itemize}
            \item \textbf{Training (Unpaired):} The $50$ degraded images form the target domain, with GTs explicitly discarded to ensure unpaired adaptation.
            \item \textbf{Testing (Paired):} Held-out pairs are reserved exclusively for full-reference quantitative evaluation.
        \end{itemize}
\end{itemize}

\paragraph{Target Domain: \texttt{Realworld-Compound} (Unpaired).}
This domain serves to adapt the model to uncontrolled, realistic scenarios and evaluate its generalization capabilities.
\begin{itemize}
    \item \textbf{Acquisition:} Diverse real-world scenes featuring natural, complex compound degradation are captured using independent optical systems.
    \item \textbf{Data Split:} No GT is available for these images. The dataset comprises $50$ unpaired training images and $51$ test images for the SL system ($50$ training, $11$ test for MRL).
    \item \textbf{Usage:} Unpaired training images are used for adaptation; test images are used for qualitative comparison and quantitative evaluation via no-reference IQA metrics.
\end{itemize}

\section{Motivation Analysis}
\label{sec:supp_motivation}
Our framework addresses compound degradation arising from optical aberrations and veiling glare.
Although we aim to mitigate both artifacts, our methodology prioritizes the modeling of the veiling glare component.
This section outlines the rationale for this focus.

\noindent\textbf{Baseline for Aberration Correction.} 
Existing literature demonstrates that supervised networks trained on paired datasets effectively handle spatially varying aberrations~\cite{chen2021optical,zhang2025degradation}.
We follow this data-driven strategy by constructing a paired source domain to train the backbone network. 
This approach yields robust performance for aberration correction.
Consequently, our work focuses on the removal of veiling glare, a more challenging and under-constrained problem, rather than the incremental improvements of aberration correction.

\noindent\textbf{Data Scarcity in Veiling Glare.}
Unlike optical aberrations, acquiring paired data for real-world veiling glare is infeasible due to the sensitivity to dynamic lighting conditions. 
Moreover, high-fidelity simulation is restricted by the inaccessibility of proprietary opto-mechanical structures and the prohibitive computational cost of non-sequential ray tracing.
This lack of GT results in an under-constrained degradation model (Eq.~\ref{eq:Img Forward} in the main text).
While the source domain constrains the PSF term $K^p$, the transmission map $T^p$ and the glare map $I_g^p$ remain unsupervised. 
Consequently, general domain adaptation methods yield suboptimal performance without explicit physical guidance.

\noindent\textbf{Targeted Design for Glare Modeling and Removal.}
To bridge the supervision gap, we introduce VeilGen for data synthesis and DeVeiler for image restoration.
Both components leverage the proposed LOTGMP prior to ensure physical consistency.
In VeilGen, the prior guides the generation of plausible transmission and glare maps in latent space. 
This synthesis is validated in two ways: the resulting degradations are more realistic (Fig.~6 in the main text), and a network trained on this synthetic data achieves superior restoration performance (Tab.~3 in the main text).
Similarly, DeVeiler incorporates the prior during restoration.
The VG-Enc module estimates latent glare and transmission, which the VGCM module then uses to selectively suppress activations in glare-affected regions, a process confirmed by our feature analysis (Fig.~8 in the main text). 
This dual application of the prior effectively addresses the challenges of data scarcity and model ambiguity.
Future work could explore a unified architecture that jointly learns to correct both aberration and glare.

\section{Generalization Analysis}

\label{sec:generalization}

While retraining a model for a specific new lens is a common practice in computational aberration correction~\cite{chen2025physics,zhang2025degradation}, our framework features a highly streamlined workflow that enables rapid and cost-effective generalization to novel optical systems. The generalization capability of our pipeline is supported by three key aspects:

\noindent\textbf{Flexible Source Data Construction.} The acquisition of source domain data in our framework is highly adaptable. If the optical design parameters are known, paired source data can be efficiently generated via optical simulation. Conversely, in strictly blind scenarios where lens parameters are completely unknown, the source data can be easily acquired through a simple screen capture setup, eliminating the need for complex calibration.

\noindent\textbf{High Target Data Efficiency.} Bridging the domain gap to a new lens typically requires extensive target data. However, our unsupervised adaptation process is remarkably efficient. As detailed in our subsequent data efficiency analysis (Fig.~\ref{fig:data_efficiency}), our model requires only $\sim$25 unpaired degraded images from the target system to achieve stable and high-quality restoration. This drastically reduces the real-world deployment cost.

\noindent\textbf{Empirical Validation on Diverse Optics.} We have rigorously validated this streamlined pipeline on two fundamentally distinct hardware prototypes: the large-aperture single lens (SL) and the metasurface-refractive hybrid lens (MRL). Notably, both evaluations were conducted under a completely blind setting, confirming that our framework is robust and readily applicable to diverse, unknown optical systems in practice.

\section{More Implementation Details}
\label{sec:Implementation Details}
\noindent\textbf{VeilGen Training.} 
In stage \Rmnum{1}, the VeilGen is initialized from the pre-trained Stable Diffusion v2-1~\cite{rombach2022high}. 
We employ AdamW~\cite{Ilya2019adamw} with a constant learning rate of $1{\times}10^{-5}$ for both the main diffusion backbone and the LOTGMP module. 
LOTGMP utilizes a ResBlock-Attention backbone, with Trans/Glare Heads implemented as convolutional layers. 
Physical consistency is enforced explicitly by Eq.~1, which constrains the prediction of Trans-Head to be multiplicative (modulation) and the Glare-Head to be additive (bias). 
This structural prior compels the heads to disentangle distinct physical roles
The model is trained for $9k$ iterations with a batch size of $16$. 
Following standard practices in hybrid-domain diffusion learning~\cite{wang2025learning}, we set the probability parameter $p$ to $0.3$ during training and the mixture coefficient $w$ to $0.85$ during inference. 
We set the reference timestep to $t^*=0$ to extract latent maps from the fully denoised state, ensuring estimation precision. 
To balance generation quality and efficiency, we employ a $10$-step sampling strategy. 
Using the Flickr2K~\cite{timofte2017ntire} dataset as the source, we synthesize $500$ high-resolution images ($1280 \times 1920$). 
This process requires approximately $3$ hours on a single NVIDIA A100 GPU.

\noindent\textbf{Distillation to DDN.} 
In stage \Rmnum{2}, we distill the trained VeilGen for $25k$ iterations using the Adam optimizer~\cite{kingma2014adam} with a cosine annealing learning rate schedule, starting from $2{\times}10^{-4}$ and decaying to $1{\times}10^{-7}$. 
The batch size is $8$, and the input patch size is $256{\times}256$. 
This distillation preserves VeilGen’s physically grounded degradation behavior while producing an efficient forward model suitable for supervision in Stage~\Rmnum{3}. 

\noindent\textbf{DeVeiler Training.} 
In stage~\Rmnum{3}, DeVeiler is trained end-to-end in two phases. 
In phase \Rmnum{1} (pre-training), we train on paired source-domain data for $100k$ iterations using Adam~\cite{kingma2014adam}, with a learning rate that decays from $2{\times}10^{-4}$ to $1{\times}10^{-7}$ via cosine annealing. 
In phase \Rmnum{2} (fine-tuning), we adapt the model on a mixed dataset of $500$ generated pseudo-pairs and the original source pairs for $5k$ iterations. 
The learning rate decays from $5{\times}10^{-5}$ to $1{\times}10^{-7}$ with cosine annealing. 
Both phases use a batch size of $8$, a $256{\times}256$ patch size, and random horizontal/vertical flips for augmentation. 
The loss weight $\lambda_{\text{rev}}$ is set to $1.0$. 
The fine-tuning phase is highly efficient, completing in approximately $40$ minutes on an NVIDIA A100 GPU. 

\noindent\textbf{Text prompts.} We utilize distinct text prompts for the source and target domains. 
The source domain prompt focuses solely on aberration: \textit{a photograph with spatial-varying PSF blur, optical aberrations, defocus, and chromatic fringing.} 
For the target domains, which include the \texttt{Screen-Compound} and \texttt{Realworld-Compound}, the prompt is expanded to describe the full compound degradation: \textit{a photograph with spatial-varying PSF blur, optical aberrations, defocus, chromatic fringing, and noticeable stray light with veiling glare.}

\noindent\textbf{Baseline Implementations.} 
For fair comparison, we use official implementations, retraining learning-based models on our dataset with default settings.

\section{Discussion}
\label{sec:discussion}
\subsection{Societal Impacts}
This work alleviates the inherent trade-off between optical miniaturization and image quality. 
While aberration correction is established, veiling glare remains a bottleneck for compact optical systems.
By addressing this compound degradation, we facilitate high-performance imaging in hardware with strict spatial limits, such as medical endoscopes, autonomous drones, and mobile devices. 
Beyond specific applications, our framework demonstrates a strategy for inverse problems in data-scarce domains. 
Instead of purely black-box approaches, we incorporate physical models into the latent space to synthesize realistic training pairs from unpaired data. 
This physics-aware generation strategy offers a scalable solution for other fields lacking GT, including underwater imaging and astronomy.

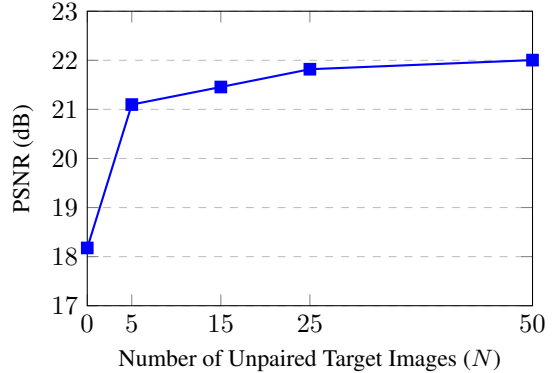
\begin{figure}[t]
\centering
\begin{tikzpicture}
\begin{axis}[
    width=0.9\linewidth,   
    height=5.5cm,          
    xlabel={Number of Unpaired Target Images ($N$)},
    ylabel={PSNR (dB)},
    xmin=0, xmax=50,
    ymin=17, ymax=23,      
    xtick={0, 5, 15, 25, 50}, 
    ytick={17, 18, 19, 20, 21, 22, 23},
    legend pos=south east, 
    ymajorgrids=true,      
    grid style=dashed,     
    xlabel style={font=\small},
    ylabel style={font=\small},
]

\addplot[
    color=blue,        
    mark=square*,      
    thick,             
    mark size=2pt      
]
coordinates {
    (0, 18.1769)
    (5, 21.0981)
    (15, 21.4561)
    (25, 21.8182)
    (50, 22.0041)
};

\end{axis}
\end{tikzpicture}
\caption{\textbf{Data Efficiency Analysis.} PSNR performance on the \texttt{Screen-Compound} test set with varying numbers of unpaired target training images ($N$). Note that $N=0$ denotes the non-adapted Source-only baseline.}
\label{fig:data_efficiency}
\end{figure}

\subsection{Limitations and Future Work}
\noindent\textbf{Limitations.} 
Although our framework delivers favorable restoration results, slight color deviations may persist under intense veiling glare (see Fig.~\ref{fig:visual_screen_mr_supp}). 
This is fundamentally due to the ambiguity in overexposed regions, where the additive glare merges with the saturated scene signal.
Regarding data acquisition, the source domain is constructed using images with a fixed scene depth.
Although aberrations are theoretically depth-dependent, real-world performance indicates that this approximation is acceptable for compact optical systems.
From a modeling perspective, the current framework learns implicit light representations rather than explicit parameters (\eg, 3D position). 
While sufficient for effective restoration, explicit parameterization could offer additional controllability in future iterations.

\noindent\textbf{Future Work.} 
To address these limitations, future research could incorporate explicit light-source modeling for color recovery and extend the source domain to include multiple scene depths. 
Additionally, we plan to investigate frequency-domain strategies for improved glare separation and adapt advanced generative backbones for efficient one-step synthesis.

\section{More Results}
\label{sec:More Results}

\noindent\textbf{Analysis of Target Data Efficiency}
\label{sec:Data Efficiency}
We analyze the impact of target training set size ($N \in \{0, 5, 15, 25, 50\}$) on restoration performance. 
To decouple the adaptation strategy from the specific architecture of DeVeiler, we employ a standard SwinIR~\cite{liang2021swinir} backbone.
Figure~\ref{fig:data_efficiency} indicates that a subset of $N=5$ yields substantial improvements over the baseline ($N=0$). 
Furthermore, performance converges at $N=15$, reaching parity with the full dataset ($N=50$). 
These results verify the capability of the framework for robust few-shot adaptation with limited data.

\begin{figure*}[!t]
  \centering
  \includegraphics[width=0.99\linewidth]{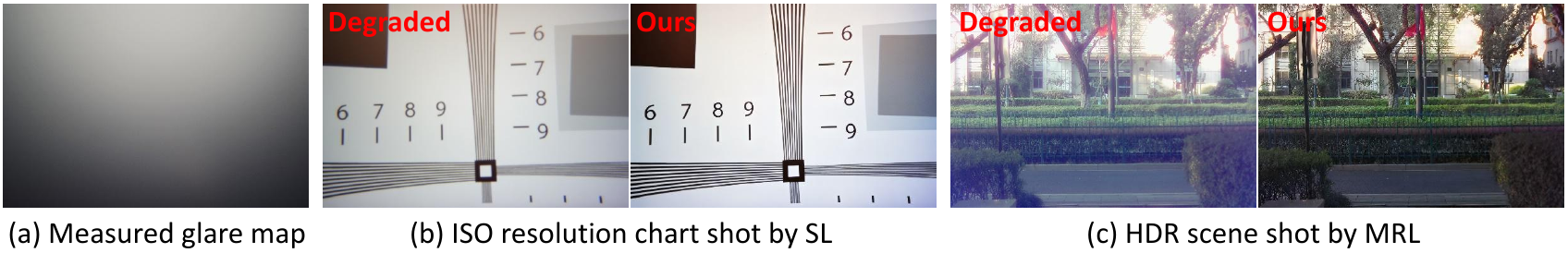}
    \caption{(a) Physically measured glare map ($I_c \approx 0$). (b) High-frequency detail recovery on an ISO resolution chart (SL). (c) Robust veiling glare removal in a high-contrast HDR environment (MRL).}
  \label{fig:visual_Challenging Scenes}
\end{figure*}

\noindent\textbf{Physical Interpretability of LOTGMP.}
We validated this via a ``black screen'' experiment (capturing $I_c \approx 0$ with side-illumination to isolate pure glare). The predicted glare map (Fig.~8(c)) exhibits strong structural consistency with physical measurements (Fig.~\ref{fig:visual_Challenging Scenes}(a)), confirming that LOTGMP accurately learns the glare distribution.

\noindent\textbf{Robustness in Challenging Scenes.}
We further demonstrate the robustness of our method on specific challenging scenarios. As shown in Fig.~\ref{fig:visual_Challenging Scenes}(b), our method effectively recovers high-frequency details in the line pairs scene captured by SL. Furthermore, it successfully removes veiling glare and preserves underlying structures in the high-contrast HDR scene captured by MRL (Fig.~\ref{fig:visual_Challenging Scenes}(c)).

\begin{figure*}[!t]
  \centering
  \includegraphics[width=0.99\linewidth]{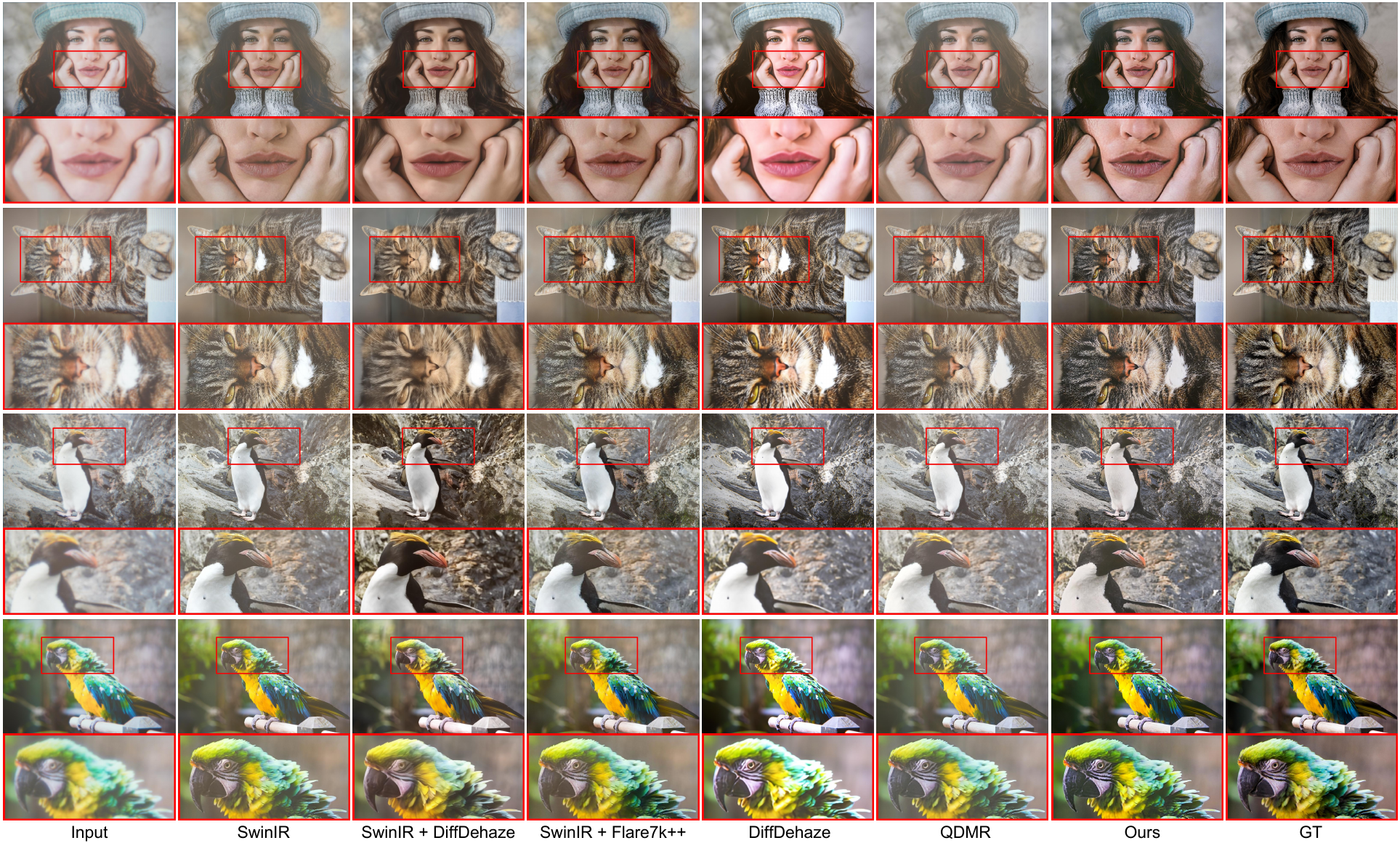}
  \caption{\textbf{Visual results on the \texttt{Screen-Compound} domain captured by the SL system.} The method is shown at the bottom of each case. Zoom in for the best view.}
  \label{fig:visual_screen_sl_supp}
\end{figure*}

\begin{figure*}[!t]
  \centering
  \includegraphics[width=0.99\linewidth]{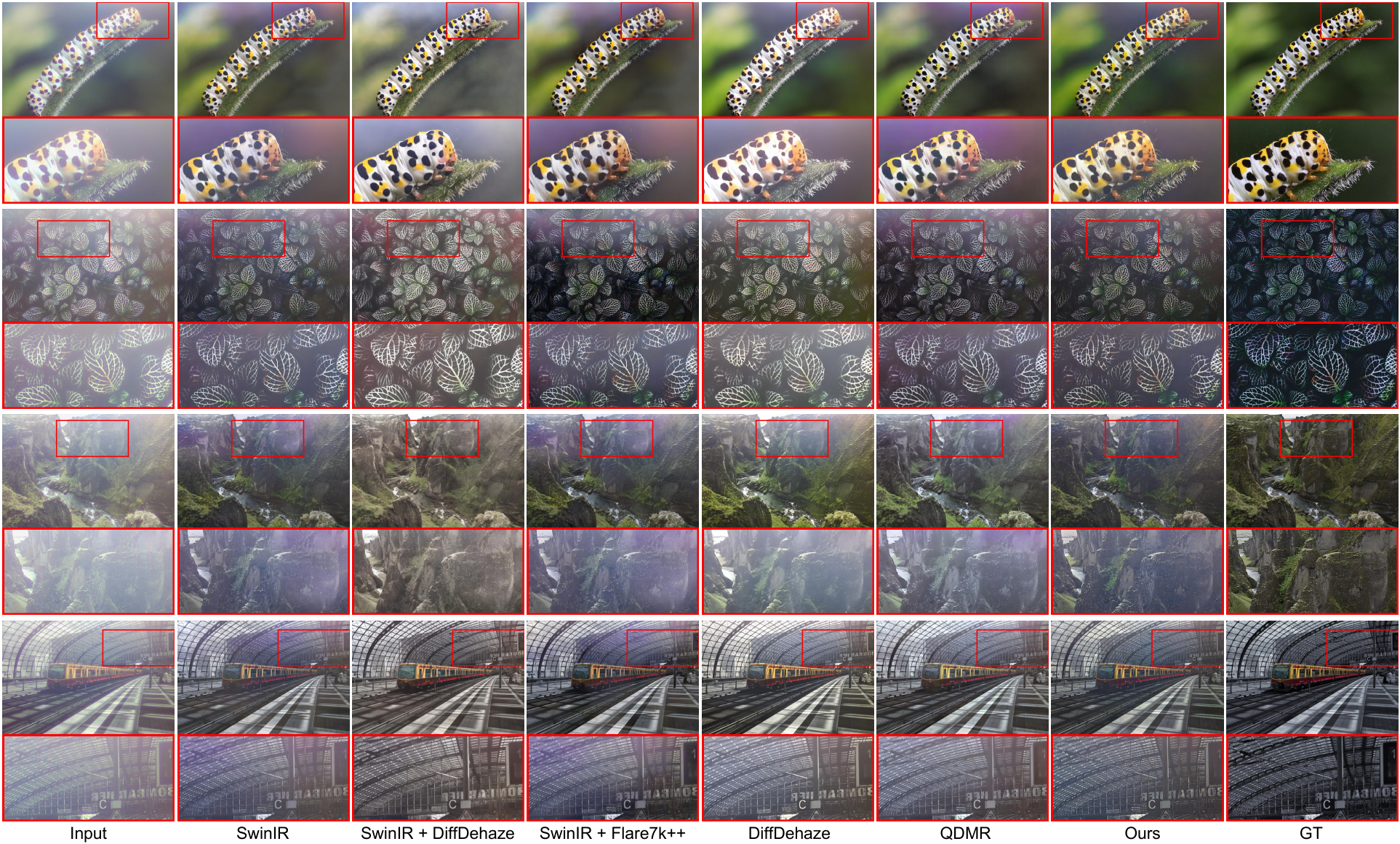}
  \caption{\textbf{Visual results on the \texttt{Screen-Compound} domain captured by the MRL system.} The method is shown at the bottom of each case. Zoom in for the best view.}
  \label{fig:visual_screen_mr_supp}

\end{figure*}

\begin{figure*}[!t]
  \centering
  \includegraphics[width=0.99\linewidth]{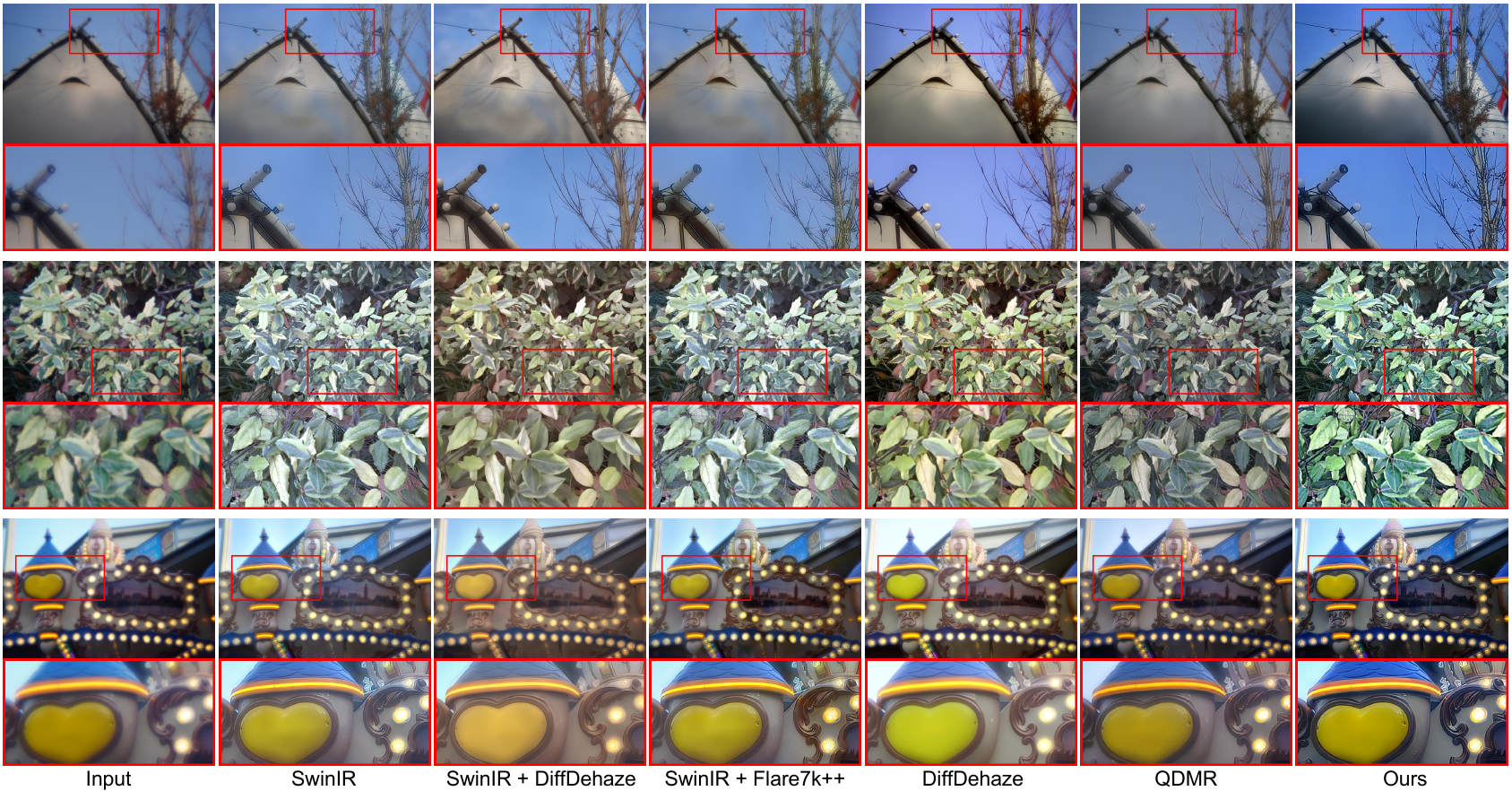}
  \caption{\textbf{Visual results on \texttt{Realworld-Compound} domain captured by the SL system.} The method is shown at the bottom of each case. Zoom in for the best view.}
  \label{fig:visual_realworld_sl_supp}
\end{figure*}

\begin{figure*}[!t]
  \centering
  \includegraphics[width=0.99\linewidth]{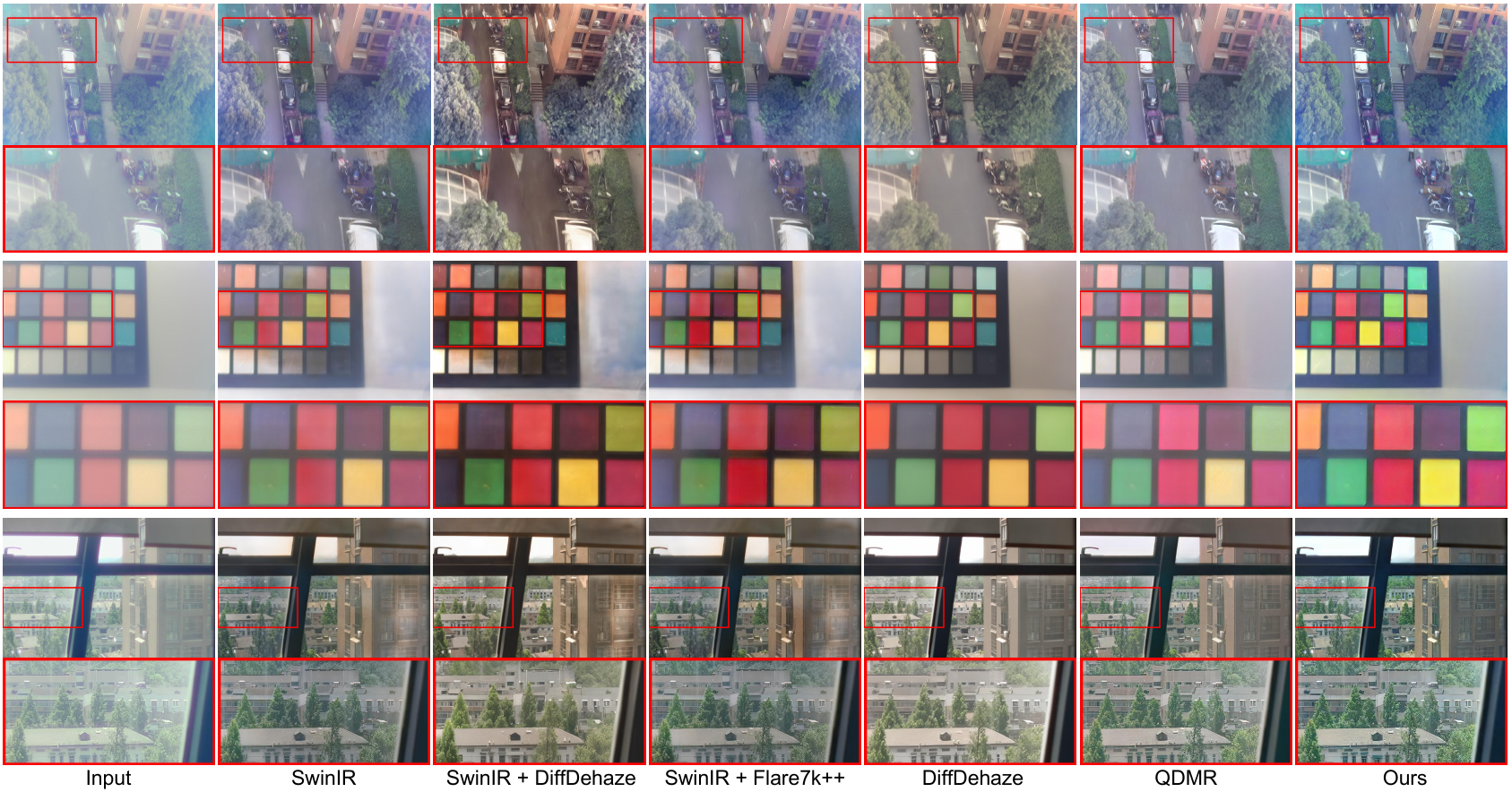}
  \caption{\textbf{Visual results on \texttt{Realworld-Compound} domain captured by the MRL system.} The method is shown at the bottom of each case. Zoom in for the best view.}
  \label{fig:visual_realworld_mr_supp}
  \vspace{-1.5em}  
\end{figure*}

\noindent\textbf{More Visual Results.}
To further verify the effectiveness of our method, we present more visual comparison results between the proposed DeVeiler and other advanced methods across both the \texttt{Screen-Compound} and \texttt{Realworld-Compound} domains. 
Specifically, the results captured by the SL and MRL systems are shown for the \texttt{Screen-Compound} domain in Fig.~\ref{fig:visual_screen_sl_supp} and Fig.~\ref{fig:visual_screen_mr_supp}, and for the \texttt{Realworld-Compound} domain in Fig.~\ref{fig:visual_realworld_sl_supp} and Fig.~\ref{fig:visual_realworld_mr_supp}, respectively.

In these scenarios, the input images exhibit compound degradation, where light sources introduce veiling glare, leading to a noticeable reduction in contrast and color shifts alongside intrinsic optical aberrations.
Baseline aberration correction methods~\cite{liang2021swinir}, trained on the source domain, show limited generalization to the unseen veiling glare. 
Cascaded pipelines employing dehazing models~\cite{wang2025learning} can improve global contrast but may result in the smoothing of fine textures.
Similarly, cascaded flare removal approaches~\cite{dai2024flare7k++}, typically designed for localized artifacts, are less effective in addressing the spatially diffusive nature of veiling glare. 
Furthermore, general domain adaptation methods~\cite{jiang2025qdmr,wang2025learning}, without explicit physical modeling for glare, may exhibit color deviations.
In contrast, DeVeiler leverages the latent veiling glare maps to achieve favorable results across diverse scenarios, preserving structural details and recovering color fidelity. 
In regions of intense illumination, the input signal approaches saturation (Fig. \ref{fig:visual_screen_mr_supp}). 
While signal loss hinders full recovery, DeVeiler minimizes color deviations compared to competing methods.

\end{document}